\documentclass[pdflatex,sn-basic, iicol]{sn-jnl}

\usepackage{subfig}

\jyear{2022}%

\theoremstyle{thmstyleone}%
\theoremstyle{thmstyletwo}%

\theoremstyle{thmstylethree}%
\newcommand{\bs}[1]{\boldsymbol{#1}}

\newcommand{\nm}{\nonumber}
\begin{document}

\title[Article Title]{Free material optimization of thermal conductivity tensors with asymmetric components}


\author[1]{\fnm{Yuki} \sur{Sato}}

\author[1,2]{\fnm{Teppei} \sur{Deguchi}}

\author[1]{\fnm{Tsuyoshi} \sur{Nomura}}

\author*[1]{\fnm{Atsushi} \sur{Kawamoto}}\email{atskwmt@mosk.tytlabs.co.jp}

\affil[1]{\orgname{Toyota Central R\&D Labs., Inc.}, \orgaddress{\street{41-1, Yokomichi}, \city{Nagakute, Aichi}, \postcode{480-1192}, \country{Japan}}}

\affil[2]{Present address: \orgdiv{Department of Physics, Graduate School of Science}, \orgname{Nagoya University}, \orgaddress{Furo-cho, Chikusa-ku, Nagoya, Aichi, \postcode{464-8602}, \country{Japan}}}

%\affil[3]{\orgdiv{Department}, \orgname{Organization}, \orgaddress{\street{Street}, \city{City}, \postcode{610101}, \state{State}, \country{Country}}}


\abstract{Free Material Optimization (FMO), a branch of topology optimization, in which the design variables are the full constitutive tensors, can provide the most general form of the design problems.
Considering the microstructure composed of isotropic materials, the constitutive tensors are yet positive definite and symmetric.
On the other hand, it has been reported that the symmetry of this constitutive tensor can be broken in appearance by considering other physical phenomena.
In the present study, we focus on the thermal Hall effect, which is explained as the phenomena that induces the temperature gradient orthogonal to a given temperature gradient across a solid when a magnetic field is applied to the solid.
This effect makes the thermal conductivity tensor asymmetric and justifies extending the space of the constitutive tensors to be an asymmetric domain.
We propose the FMO for asymmetric constitutive tensors, parameterizing the design space so that the physically available property could be naturally satisfied.
Several numerical experiments are provided to show the validity and the utility of the proposed method.
}


\maketitle
\section{Introduction}\label{sec:intro}
The original topology optimization, first implemented on computers  by \cite{bendsoe1988generating}, was based on the homogenization method, which relaxed the design space by assuming infinitely small holes.
This approach was aware of the microstructures existing behind the macroscale behaviors.
Then, the density method~\citep{bendsoe1999material}, which does not necessarily assume microstructures, came to be preferred in terms of computational resources.
This method, on the other hand, can be viewed as an approach of directly optimizing macroscopic structures.

From the computational costs point of view, we can also consider an approach in which the features of the microstructures are embedded in macroscopic characteristics using some parameters, such as orientation angles of anisotropic materials, and are optimized.
\cite{pedersen1989optimal} dealt with the local orientation distribution based on the two-dimensional rotation of anisotropic materials.
\cite{nomura2019inverse} introduced orientation tensors as the design variables that correspond to the tensor product of the associated orientation vectors.
For ensuring that the design variables actually have the property of the orientation tensors, symmetric and trace-preserving constraints are imposed on the design variables.
\cite{jung2022inverse} proposed a multiscale topology optimization in which the spatially-varying fiber size and orientation in three-dimensional fiber reinforced composites were optimized.

Another approach with a higher degree of freedom is to optimize the macroscopic material properties themselves. 
This approach is called free material optimization (FMO).
In the original FMO conducted by \cite{zowe1997free}, the design variable is the full elastic stiffness tensor, which is physically available.
Since it offers maximum flexibility of the representation of macroscopic structures and the microscopic orientation simultaneously, FMO gives the best material configuration in the physical sense~\citep{kovcvara2008free}.

These anisotropic topology optimization methods successfully conducted multiscale optimization, but the searched space of the constitutive tensors was confined within symmetry because the target physical properties were well expressed by  the symmetric constitutive tensors.
In this study, on the other hand, we focus on the physical phenomenon, in which the symmetry of the constitutive tensors is apparently broken, i.e. the thermal Hall effect.

The thermal Hall effect is explained as the phenomenon that induces the temperature gradient orthogonal to a given temperature gradient across a solid when a magnetic field is applied to the solid~\citep{ben2016photon, murakami2017thermal}.
This effect makes the thermal conductivity tensor asymmetric and justifies extending the space of the constitutive tensors to be an asymmetric domain.
Enlarging the design space from the symmetric constitutive tensors to the asymmetric ones, the theoretical performance limit of devices would be improved.

The present study, therefore, proposes FMO for asymmetric thermal conductivity tensors whose asymmetry comes from the magnetic field to be applied to the material.
That is, free material optimization for asymmetric thermal conductivity tensors is the simultaneous optimization of topology, orientation directions, and the external magnetic field.

The remainder of this paper is organized as follows.
Section~\ref{sec:methods} describes the proposed method.
First, we briefly explain the thermal Hall effect, followed by the derivation of the governing equation.
Then, the optimization problems are formulated, and the optimization algorithm we used in this study is described.
In Section~\ref{sec:experiments}, several numerical experiments are provided to demonstrate the utility of the proposed method.
Finally, we conclude this study in Section~\ref{sec:conclusion}.

\section{Methods} \label{sec:methods}
\subsection{Thermal Hall effect} \label{subsec:thermal}
The thermal Hall effect is the thermal analog of the Hall effect and was independently discovered by Righi and Leduc.
The thermal Hall effect is the phenomenon that induces the temperature gradient orthogonal to a given temperature gradient when a magnetic field is applied to particular materials.

Let $T$ be the temperature and $B_z$ be the magnetic field along the $z$-axis.
Given the temperature gradient introduced along the $x$-axis, the thermal Hall effect induces the temperature gradient along the $y$-axis in the presence of the magnetic field $B_z$.
The ratio of the temperature gradients is defined as a thermal Hall coefficient $R^\text{TH}$~\citep{kobayashi2012large} as follows:
\begin{align}
    R^\text{TH} := \dfrac{1}{B_z}\dfrac{\frac{\partial T}{\partial y}}{\frac{\partial T}{\partial x}}. \label{eq:TH_def0}
\end{align}
This coefficient depends on the material, the temperature and the magnitude of the magnetic field.

In the present study, we model the thermal Hall effect as the expansion of Eq.~\eqref{eq:TH_def0}:
\begin{align}
    \bs{q}^\text{TH} = - R^\text{TH} \bs{B} \times k \nabla T, \label{eq:TH_def}
\end{align}
where $\bs{q}^\text{TH}$ is the heat flux induced by the thermal Hall effect, $k$ is the thermal conductivity, and $\bs{B}$ is the magnetic flux.
Then, the total heat flux in a solid is given as
\begin{align}
    \bs{q} &= -k \nabla T + \bs{q}^\text{TH} \nm \\
    &= -k \nabla T - R^\text{TH} \bs{B} \times k \nabla T.
\end{align}
In two-dimensional settings, where the magnetic flux is along $z$-axis and is denoted by $B_z$, this total heat flux is given as
\begin{align}
    \bs{q} &= - \left(k\bs{I} + \bs{k}^\text{TH} \right) \nabla T,
\end{align}
where $\bs{I}$ is the identity matrix and $\bs{k}^\text{TH}$ is the fictitious thermal conductivity tensor defined as
\begin{align}
    \bs{k}^\text{TH} :=  R^\text{TH} B_z \begin{pmatrix}
    0 & - k \\
    k & 0
    \end{pmatrix}.
\end{align}
Due to the existence of $\bs{k}^\text{TH}$, the effective thermal conductivity tensor thus becomes asymmetric.

In the present study, we expand this effective thermal conductivity tensor to a composite material, which is composed of an isotropic material exhibiting the thermal Hall effect and an anisotropic material not exhibiting it.
We define the thermal conductivity tensor of the anisotropic material as
\begin{align}
    \bs{k}^\text{aniso} := \begin{pmatrix}
    k_{11} & k_{12} \\
    k_{12} & k_{22}
    \end{pmatrix},
\end{align}
where $k_{11}$, $k_{12}$, and $k_{22}$ are the components of the thermal conductivity tensor of the anisotropic material.
Then, the effective thermal conductivity tensor is defined as follows:
\begin{align}
    \bs{k} :=& k\bs{I} + \bs{k}^\text{aniso} + \bs{k}^\text{TH} \nm \\
    =& \begin{pmatrix}
    k + k_{11} & k_{12} - R^\text{TH} B_z k \\
    k_{12} + R^\text{TH} B_z k & k + k_{22}
    \end{pmatrix}.
\end{align}
This tensor is anisotropic and asymmetric.
The present study optimize the anisotropic material $\bs{k}^\text{aniso}$ and the magnetic field $B_z$ based on the concept of FMO.

\subsection{Governing equation} \label{subsec:gov}
Consider an open bounded set $\Omega \subset \mathbb{R}^d$ where $d$ is the spatial dimension.
Let $T: \Omega \rightarrow \mathbb{R}$ and $Q: \Omega \rightarrow \mathbb{R}$ respectively denote the temperature and heat source in the domain $\Omega$.
We now assume the steady-state heat conduction in the domain $\Omega$, which is occupied by a material with the thermal conductivity tensor $\bs{k}$.
The temperature in the domain $\Omega$ is then governed by
%%%%%%% equation %%%%%%%%%%%%%%
\begin{align}
    -\nabla \cdot \left( \bs{k} \nabla T \right) = Q \quad \text{in}~\Omega. \label{eq:gov}
\end{align}
%%%%%%% equation %%%%%%%%%%%%%%
As the boundary condition, we assume that the Neumann and Dirichlet boundary conditions are respectively imposed on the boundaries $\Gamma_\mathrm{N}$ and $\Gamma_\mathrm{D}$, where $\Gamma_\mathrm{N} \cup \Gamma_\mathrm{D} = \partial \Omega$ as follows:
%%%%%%% equation %%%%%%%%%%%%%%
\begin{align}
\begin{cases}
    -\bs{n} \cdot \left( \bs{k} \nabla T \right) = 0 & \text{on}~\Gamma_\mathrm{N} \\
    T = 0 & \mathrm{on}~\Gamma_\mathrm{D}.
\end{cases} \label{eq:bnd}
\end{align}
%%%%%%% equation %%%%%%%%%%%%%%
For applying the finite element method, the weak form of the governing equation in Eq.~(\ref{eq:gov}) with the boundary conditions in Eq.~(\ref{eq:bnd}) is derived as 
%%%%%%% equation %%%%%%%%%%%%%%
\begin{align}
    \int_\Omega \tilde{T} Q ~d\Omega - \int_\Omega \nabla \tilde{T} \cdot \left( \bs{k} \nabla T \right) ~d\Omega = 0 \nm \\
    \text{for}~T, \forall \tilde{T} \in \mathcal{T}, \label{eq:gov_weak}
\end{align}
%%%%%%% equation %%%%%%%%%%%%%%
where $\mathcal{T}$ is defined as 
%%%%%%% equation %%%%%%%%%%%%%%
\begin{align}
    \mathcal{T} := \left\{ T \in H^1(\Omega) \mid T = 0 ~\text{on}~ \Gamma_\mathrm{D} \right\}, \label{eq:T_space}
\end{align}
%%%%%%% equation %%%%%%%%%%%%%%
where $H^1(\Omega)$ is the Sobolev space.
To ensure that the governing equation has the unique solution, the symmetric part of the effective thermal conductivity, i.e., $(\bs{k} + \bs{k}^\text{T})/2$ must be positive-definite based on the Lax-Milgram theorem.

\subsection{Optimization problem}
\subsubsection{Design variables}
Here, we derive design variables representing the effective thermal conductivity tensors which satisfy the Lax-Milgram theorem, that is, which are physically available.
First, to ensure the positiveness of $\bs{k}^\text{aniso}$, the following conditions are required:
%%%%%%% equation %%%%%%%%%%%%%%
\begin{align}
\begin{cases}
    & \mathrm{tr} (\bs{k}^\text{aniso}) = k_{11} + k_{22} > 0 \\
    & \mathrm{det} (\bs{k}^\text{aniso}) = k_{11}k_{22} - k_{12}^2 > 0.
\end{cases} \label{eq:pw_positive}
\end{align}
%%%%%%% equation %%%%%%%%%%%%%%
Under the satisfaction of these conditions, the symmetric part of the effective thermal conductivity $(\bs{k} + \bs{k}^\text{T})/2$ is positive-definite as follows:
\begin{align}
\begin{cases}
    & \mathrm{tr} \left( \dfrac{\bs{k} + \bs{k}^\text{T}}{2} \right) \\
    &\quad = 2k + \mathrm{tr}(\bs{k}^\text{aniso}) > 0 \\
    & \mathrm{det} \left( \dfrac{\bs{k} + \bs{k}^\text{T}}{2} \right) \\
    &\quad = k^2 + k \mathrm{tr}(\bs{k}^\text{aniso}) + \mathrm{det} (\bs{k}^\text{aniso}) > 0.
\end{cases}
\end{align}
Therefore, it is sufficient to consider the conditions \eqref{eq:pw_positive} for ensuring the existence of the solution of the governing equation.

Second, to constrain the allowable amount of material, the upper bound is set to the trace of the thermal conductivity tensor as
%%%%%%% equation %%%%%%%%%%%%%%
\begin{align}
    \mathrm{tr} (\bs{k}^\text{aniso} ) = k_{11} + k_{22} \leq c, \label{eq:pw_trace}
\end{align}
%%%%%%% equation %%%%%%%%%%%%%%
where $c$ is a positive constant.

Third, we impose the bound on the component of the asymmetric part of $\bs{k}$ as follows:
%%%%%%% equation %%%%%%%%%%%%%%
\begin{align}
   \lvert  R^\text{TH}B_z \rvert \leq b, \label{eq:pw_asym}
\end{align}
%%%%%%% equation %%%%%%%%%%%%%%
where $b$ is a positive parameter set by experimental results.

Now, we parameterize the effective thermal conductivity tensor by introducing the design variables.
Let $\xi \in L^\infty (\Omega; [-1, 1 ])$ and $\eta \in L^\infty (\Omega; [-1, 1 ])$ be the design variable fields and are mapped onto the space of diagonal components of $\bs{k}^\text{aniso}$ using the shape function as follows:
\begin{align}
    \begin{pmatrix} k_{11} \\ k_{22} \end{pmatrix} = \sum_{i=1}^4 N_i (\xi, \eta) \bs{v}_i
\end{align}
where $N_i$ and $\bs{v}_i$ are respectively the shape function and the coordinates defined as
\begin{align}
    N_1 &= \dfrac{1}{2} \left( 1 - \xi \right) \left( 1 - \xi \right)  \\
    N_2 &= \dfrac{1}{2} \left( 1 + \xi \right) \left( 1 - \xi \right) \\
    N_3 &= \dfrac{1}{2} \left( 1 - \xi \right) \left( 1 + \xi \right)  \\
    N_4 &= \dfrac{1}{2} \left( 1 + \xi \right) \left( 1 + \xi \right),
\end{align}
and 
\begin{align}
    \bs{v}_1 &= \left( c\varepsilon/2, c\varepsilon/2 \right)^\top \\
    \bs{v}_2 &= \left( c - c\varepsilon/2, c\varepsilon/2 \right)^\top \\
    \bs{v}_3 &= \left( c\varepsilon/2, c - c\varepsilon/2 \right)^\top \\
    \bs{v}_4 &= \left( c/2, c/2 \right)^\top.
\end{align}
$\varepsilon$ is a small constant for avoiding that the trace of $\bs{k}^\text{aniso}$ becomes $0$.
This mapping ensures that the trace of $\bs{k}^\text{aniso}$ is larger than or equal to $c\varepsilon$.

Next, to ensure the positiveness of the determinant of $\bs{k}^\text{aniso}$, we represent the off-diagonal component of $\bs{k}^\text{aniso}$ using the design variable field $s \in L^\infty (\Omega; [-1, 1 ])$ as 
\begin{align}
    k_{12} = s \sqrt{(1-\varepsilon')k_{11} k_{22}},
\end{align}
which naturally satisfies that the determinant is positive:
\begin{align}
    \text{det}(\bs{k}^\text{aniso}) &= k_{11} k_{22} - k_{12}^2 \nm \\
    &= k_{11} k_{22} (1-s^2(1-\varepsilon')) > 0,
\end{align}
where $\varepsilon'$ is a small constant to ensure the positiveness of the determinant.

Lastly, we parameterize the fictitious thermal conductivity tensor, using the design variable field $a \in L^\infty (\Omega; [-1, 1])$ as
\begin{align}
    a = \dfrac{R^\text{TH}B_z}{b}.
\end{align}
This definition automatically satisfies the condition in Eq.~\eqref{eq:pw_asym}

As a summary, the thermal conductivity tensor is now represented by four design variable fields, $\xi, \eta, s, a$, as
\begin{align}
    \bs{k} = \begin{bmatrix}
        k + k_{11}(\xi, \eta) & k_{12}(s) - ab k \\
        k^\text{mat}_{12}(s) + ab k & k + k_{22}(\xi, \eta)
    \end{bmatrix}. \label{eq:k_param}
\end{align}

\subsubsection{Objective functional} \label{sec:objective}
In the present study, the objective is to find an optimal material property that realizes the desired heat manipulation.
We consider two kinds of problems; the temperature minimization problem and the heat path switching problem.

\paragraph{Temperature minimization problem}
Let $\Omega_\mathrm{p} \subset \Omega$ denote the domain which should be protected from being high temperature.
The optimization problem is then formulated as follows:
\begin{align}
    & \min_{\xi, \eta, s, a} \int_{\Omega_\mathrm{p}} T ~d\Omega \\
    & \text{subject to: } \nm \\
    & \quad \int_\Omega \tilde{T} Q ~d\Omega - \int_\Omega \nabla \tilde{T} \cdot \left( \bs{k}(\xi, \eta, s, a) \nabla T \right) ~d\Omega = 0 \nm \\
    & \quad \text{for}~T, ~\forall \tilde{T} \in \mathcal{T}.
\end{align}

\paragraph{Heat path switching problem}
The symmetric part of the thermal conductivity tensor $\bs{k}$ is determined by the material while the asymmetric part of it is determined by the magnetic field applied to the material through the thermal Hall effect.
Here, we consider switching the heat path in  material depending on the magnetic field applied to the material.
Let $\Omega_\mathrm{p} \subset \Omega$ and $\Omega_\mathrm{p'} \subset \Omega$ respectively denote the domains where the temperature is minimized or maximized according to the magnetic field.
We formulate an optimization problem as follows:
\begin{align}
    & \min_{\xi, \eta, s, a, a'} \int_{\Omega_\mathrm{p}} T ~d\Omega - \int_{\Omega_\mathrm{p'}} T ~d\Omega \nm \\
    & \qquad \qquad + \int_{\Omega_\mathrm{p'}} T' ~d\Omega - \int_{\Omega_\mathrm{p}} T' ~d\Omega \\
    & \text{subject to: } \nm \\
    & \quad \int_\Omega \tilde{T} Q ~d\Omega - \int_\Omega \nabla \tilde{T} \cdot \left( \bs{k}(\xi, \eta, s, a) \nabla T \right) ~d\Omega = 0 \nm \\
    & \quad \text{for}~T, ~\forall \tilde{T} \in \mathcal{T}. \\
    & ~ \int_\Omega \tilde{T}' Q ~d\Omega - \int_\Omega \nabla \tilde{T}' \cdot \left( \bs{k}(\xi, \eta, s, a') \nabla T' \right) ~d\Omega = 0 \nm \\
    &\quad \text{for}~T', ~\forall \tilde{T}' \in \mathcal{T}.
\end{align}
The first and the second terms represent minimizing and maximizing the temperature $T$ in the domains $\Omega_\mathrm{p}$ and $\Omega_\mathrm{p'}$, respectively.
The temperature $T$ is the field realized when the asymmetric part of the thermal conductivity tensor is represented by the parameter $a$.
On the other hand, the third and the fourth terms represent minimizing and maximizing the temperature $T'$ in the domains $\Omega_\mathrm{p'}$ and $\Omega_\mathrm{p}$, respectively.
The temperature $T'$ is the response by the asymmetric part of the thermal conductivity tensor represented by the parameter $a'$.
This objective function represents the switching of heat path depending on whether the asymmetric part of the thermal conductivity tensor is determined by $a$ or $a'$.

\subsection{Optimization algorithm} \label{sec:algo}
Here, we describe the optimization algorithm used to solve the optimization problems formulated in the previous section.
Introducing the fictitious time evolution equation with the form of the reaction-diffusion equation, we update the design variable $\phi \in \{\xi, \eta, s, a \}$ as follows:
\begin{align}
    \dfrac{\partial \phi}{\partial t} = - \mathcal{L}' + R^2 \nabla^2 \phi,
\end{align}
where $\mathcal{L}'$ is the design sensitivity and $R$ is the parameter.
During the update, the second term on the right-hand side, the diffusion term, ensures the smoothness of the design variable field.
We derive the weak form of the reaction-diffusion equation, discretizing it by the finite difference method in the time direction as follows:
\begin{align}
    & \int_\Omega \tilde{\phi}(\bs{x}) \phi(t, \bs{x}) d \Omega - \int_\Omega \tilde{\phi}(\bs{x}) \phi(t-\Delta t, \bs{x}) d\Omega \nm \\
    & = - \int_\Omega \Delta t \tilde{\phi}(\bs{x}) \mathcal{L}' d\Omega \nm \\
    & \quad - \int_\Omega \Delta t R^2 \nabla \tilde{\phi}(t, \bs{x}) \cdot \nabla \phi(t, \bs{x}) d\Omega, \label{eq:rde}
\end{align}
where $\Delta t$ is the time interval for the time integration.
In the present study, the design sensitivity $\mathcal{L}'$ is given using the first and the second moment of the gradient of the objective functional based on the concept of the adaptive moment estimation (ADAM)~\citep{kingma2014adam}.
Let $G(t, \bs{x})$ be the gradient of the objective function at the (fictitious) time $t$ and the coordinate $\bs{x}$, which can be obtained by the adjoint variable method.
The first and the second moment of the gradient, respectively denoted by $v(t, \bs{x})$ and $s(t, \bs{x})$, are given as
\begin{align}
    v(t, \bs{x}) &= \beta_1 v(t-\Delta t, \bs{x}) + (1-\beta_1) G(t-\Delta t, \bs{x}) \\
    s(t, \bs{x}) &= \beta_2 s(t-\Delta t, \bs{x}) + (1-\beta_2) G(t-\Delta t, \bs{x})^2,
\end{align}
where $\beta_1$ and $\beta_2$ are the hyper-parameters.
Using these moments, the design sensitivity $\mathcal{L}'$ is defined as
\begin{align}
    \mathcal{L}' = \dfrac{v(t, \bs{x})}{\sqrt{s(t, \bs{x}) + \epsilon}}, \label{eq:sens}
\end{align}
where $\epsilon$ is the small constant for avoiding zero division.
The momentum (the first moment) is the moving average of the gradient, so it has a small value when it oscillates within some iterations.
Therefore, the use of momentum mitigates oscillations during optimization.
The second moment adaptively controls the step size, i.e., the time interval for the time integration.
When the optimization oscillates, that is, the sign of the gradient changes in every iteration, the momentum cancels while the second moment has a large value.
As a result, the definition of the design sensitivity in Eq.~\eqref{eq:sens} has a small value that leads to a small time interval.

Based on the scheme mentioned above, the optimization algorithm is constructed as follows:
\begin{enumerate}
    \item Set initial design variable fields as $\xi = 0$, $\eta = 0$, $s = 0$, and $a = 0$.
    \item Map the design variable fields onto the effective thermal conductivity tensor in Eq.~\eqref{eq:k_param}.
    \item Solve the governing equation in Eq.~\eqref{eq:gov_weak} using the finite element method.
    \item Evaluate the objective functional value. If the objective functional value converges, the optimization process is stopped. Otherwise, proceed to the next step.
    \item Solve the adjoint equation, which is derived by the adjoint variable method, by the finite element method. The design sensitivity is then calculated.
    \item Update the design variable fields using the reaction-diffusion equation in Eq.~\eqref{eq:rde}, which is solved by the finite element method.
    \item Return to the second step.
\end{enumerate}

\section{Numerical experiments} \label{sec:experiments}
\subsection{Computational conditions}
Here, we describe the computational conditions, setting parameters used in all the following examples.
The parameters for the optimizer described in Section~\ref{sec:algo} were set to $\beta_1 = 0.9$, $\beta_2 = 0.999$, $\epsilon = 1.0 \times 10^{-8}$ and $\Delta t = 1.0 \times 10^{-2}$, respectively.
The optimization processes were stopped when the following condition was satisfied:
\begin{equation}
    \dfrac{\lvert J(t) - J(t-\Delta t) \rvert}{\lvert J(t-\Delta t) \rvert} \leq 1.0 \times 10^{-6},
\end{equation}
when $J(t)$ is the objective functional value at the fictitious time $t$.

The finite element method was used to solve the governing and adjoint equations, and the reaction-diffusion equation for updating the design variable fields.
We used the \textit{scikit-fem}~\citep{skfem2020}, which is open-source software for the finite element assembler, for implementing the finite element method.

The thermal conductivity of the material, which exhibits the thermal Hall effect, $k$ was set to $10$.
The upper limit of the trace of the anisotropic material was set to $c=20$.
In all the following experiments, $Q$ in \eqref{eq:gov_weak} was set to
\begin{align}
    Q = \begin{cases}
    1.0 \times 10^{5} & \mathrm{in} ~ \Omega_\mathrm{h} \\
    0 & \mathrm{in} ~ \Omega \setminus \Omega_\mathrm{h},
    \end{cases}
\end{align}
where $\Omega_\mathrm{h} \subset \Omega$ is the domain having the heat source.
We set the parameter $b$, which bounded the upper limit of $B_z R^\text{TH}$, to $0.3$ when the asymmetric part was dealt with. 
This value was determined based on the experimental results of Bismuth, which is known as having the relatively large thermal Hall coefficient~\citep{kobayashi2012large}.

\subsection{Forward analysis}
%%%%%%% figure %%%%%%%%%%%%%%
\begin{figure}[t]
\centering
\includegraphics[width=6cm]{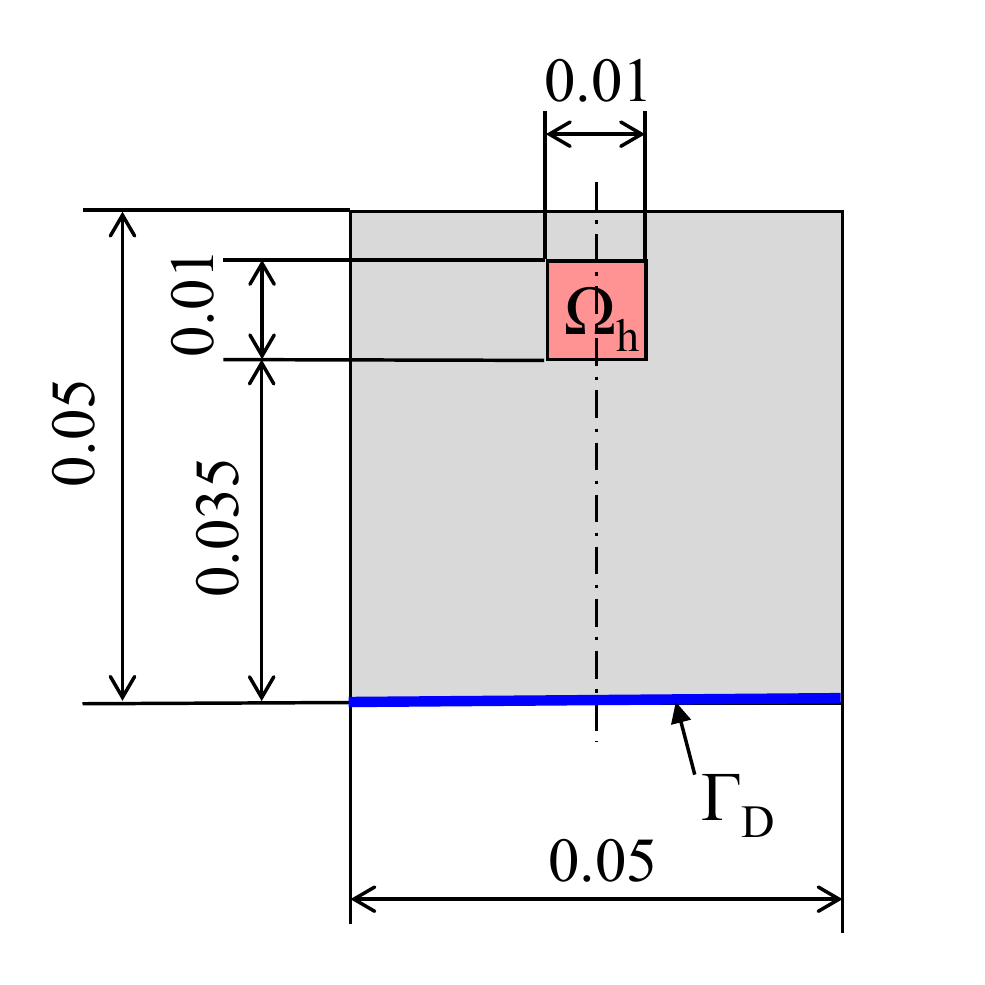}
\caption{Analysis settings. \label{fig:problem0}}
\end{figure}
%%%%%%% figure %%%%%%%%%%%%%%
%%%%%%% figure %%%%%%%%%%%%%%
\begin{figure*}[t]
\centering
\subfloat[]{\includegraphics[width=0.25\textwidth]{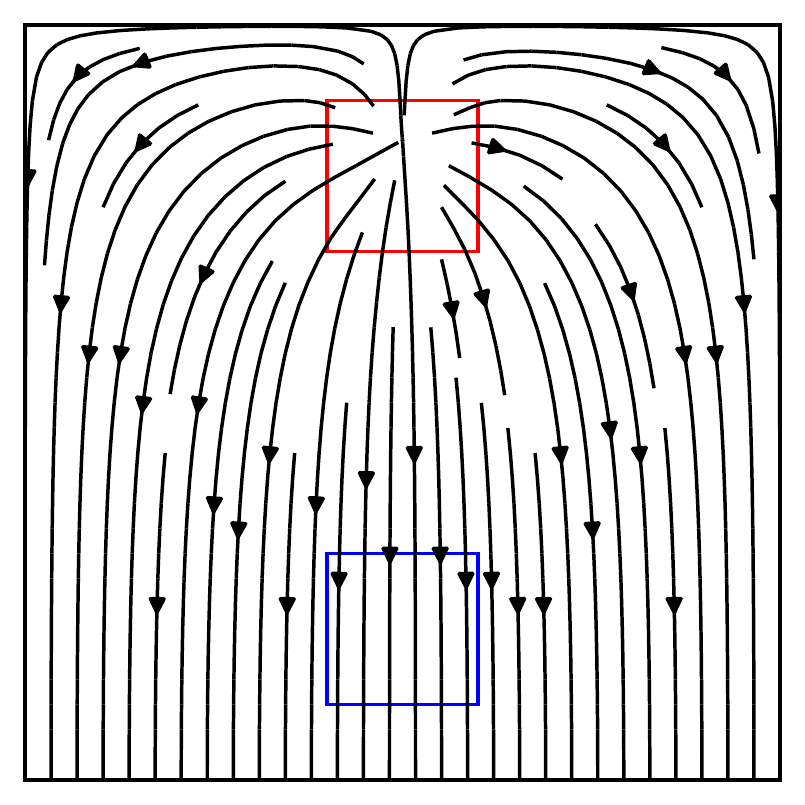} \label{fig:flux0}}
\subfloat[]{\includegraphics[width=0.25\textwidth]{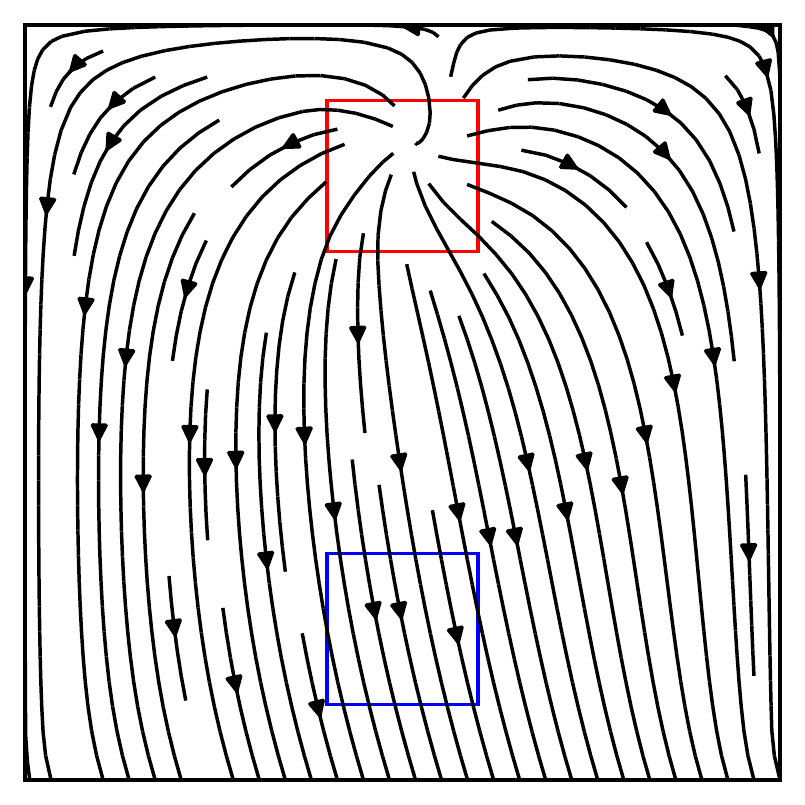} \label{fig:flux1}}
\subfloat[]{\includegraphics[width=0.25\textwidth]{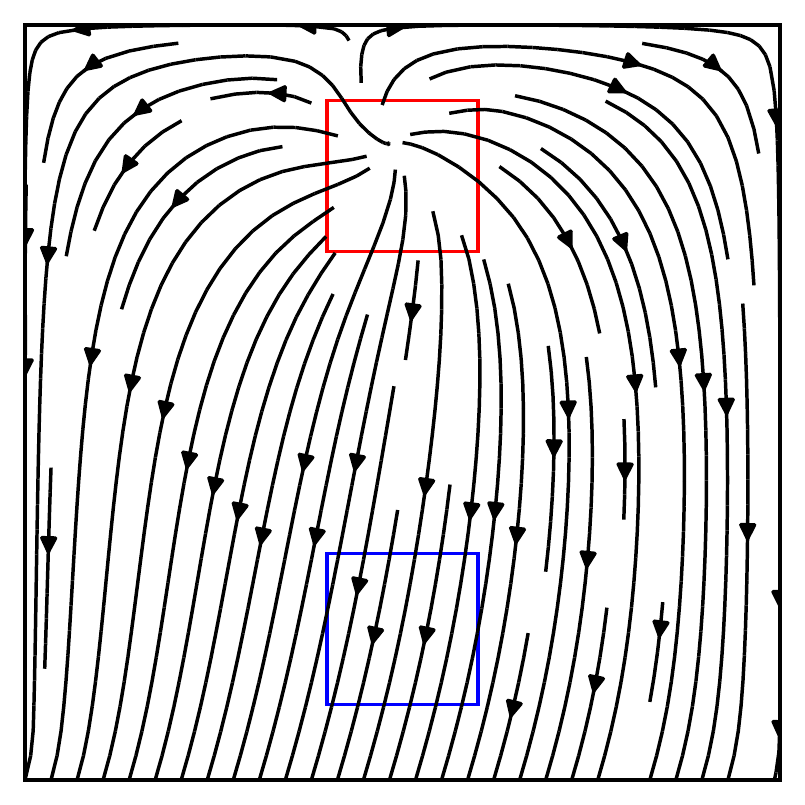} \label{fig:flux2}} \\
\subfloat[]{\includegraphics[width=0.3\textwidth]{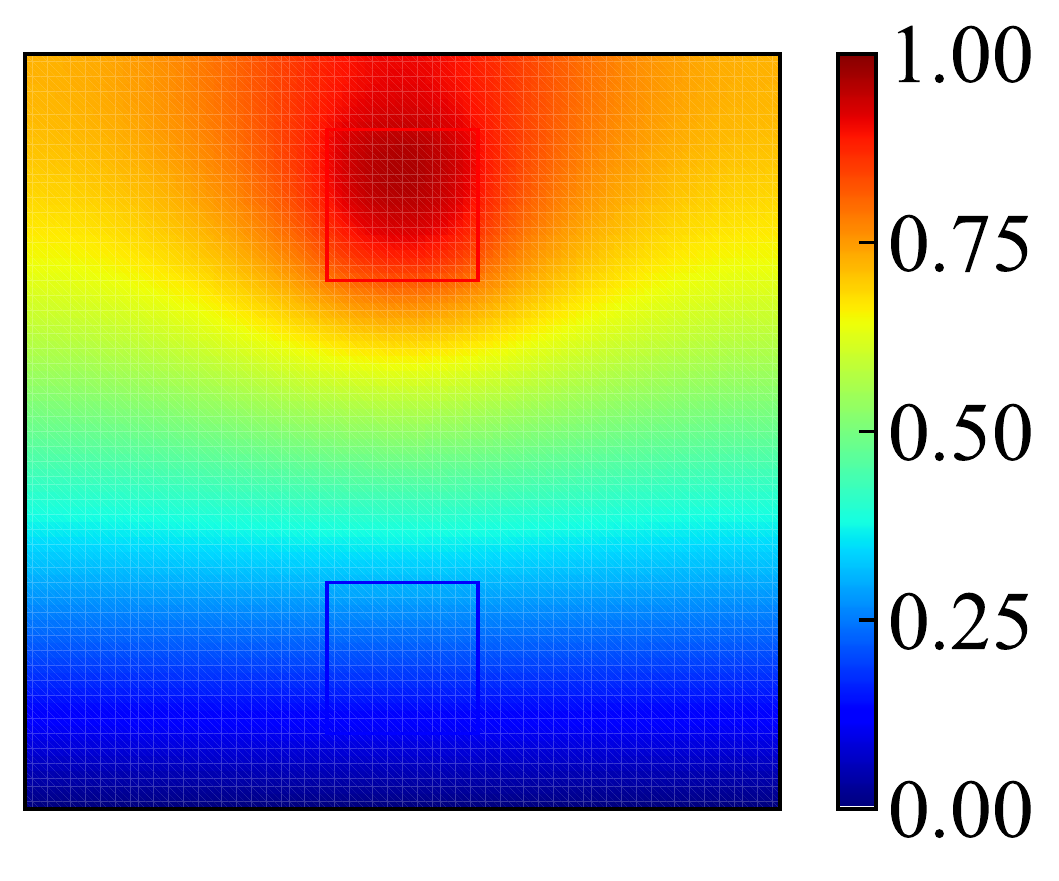} \label{fig:temp0}}
\subfloat[]{\includegraphics[width=0.3\textwidth]{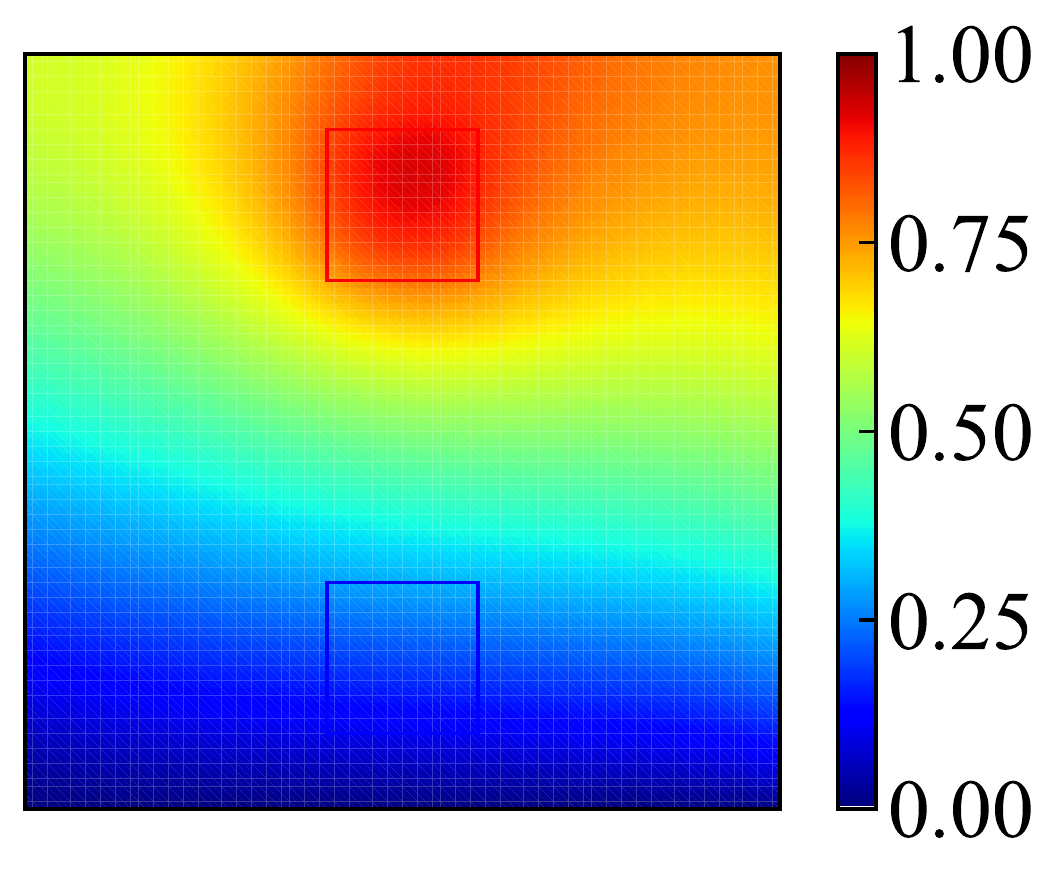} \label{fig:temp1}}
\subfloat[]{\includegraphics[width=0.3\textwidth]{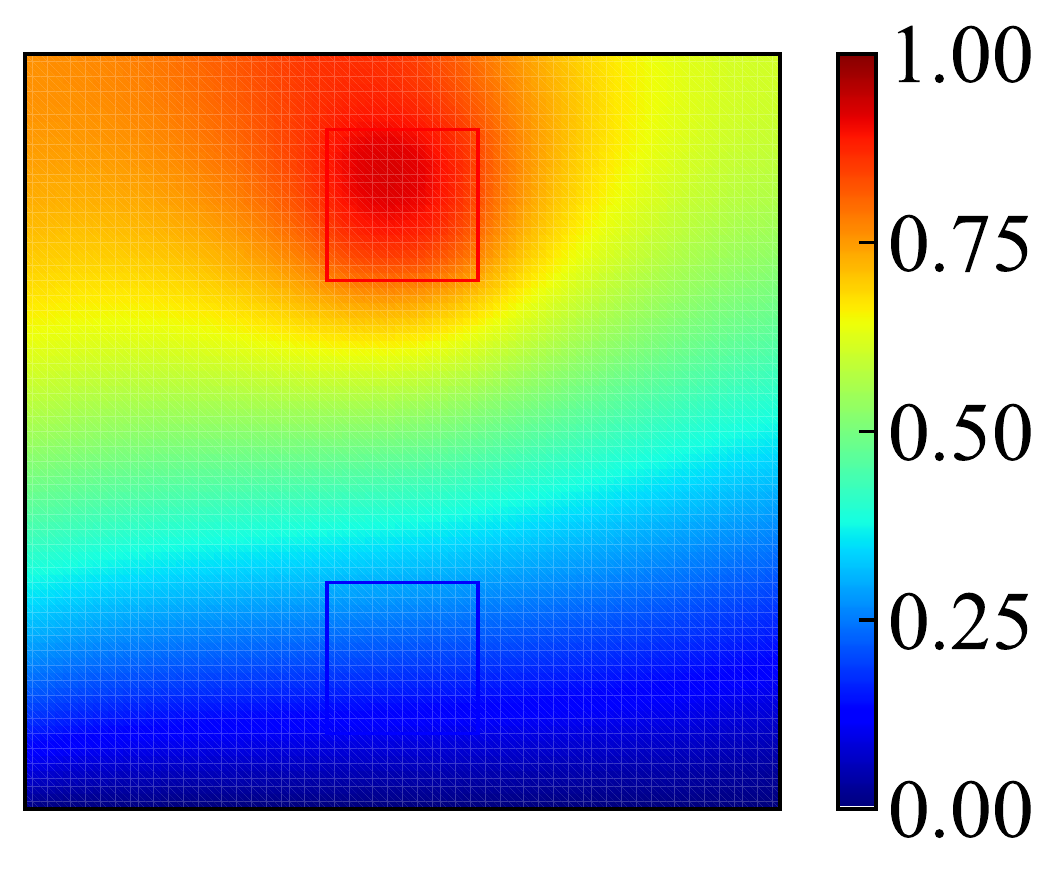} \label{fig:temp2}}
\caption{Forward analysis results. The heat fluxes were calculated for (a) Case 1-1: $\xi=-1$, $\eta=-1$, $s=0$, and $a=0$, (b) Case 1-2: $\xi=-1$, $\eta=-1$, $s=0$, and $a=1$, and (c) Case 1-3: $\xi=-1$, $\eta=-1$, $s=0$, and $a=-1$. The temperature fields for each condition are respectively shown in (d) Case 1-1: $\xi=-1$, $\eta=-1$, $s=0$, and $a=0$, (e) Case 1-2: $\xi=-1$, $\eta=-1$, $s=0$, and $a=1$, and (f) Case 1-3: $\xi=-1$, $\eta=-1$, $s=0$, and $a=-1$.\label{fig:result0} }
\end{figure*}
%%%%%%% figure %%%%%%%%%%%%%%
First, we perform the forward analysis dealing with the thermal Hall effect.
Figure~\ref{fig:problem0} illustrates the problem settings for the forward analysis.
Here, we considered the following three cases, setting parameters for ensuring the positiveness of $\bs{k}^\text{aniso}$ to $\varepsilon = 1.0 \times 10^{-4}$ and $\varepsilon' = 1.0 \times 10^{-4}$, respectively.
In Case 1-1, the design variable fields were set to $\xi=-1$, $\eta=-1$, $s=0$, and $a=0$.
This resulted in $\bs{k}^\text{aniso} = c \varepsilon \bs{I}$, which means that the total thermal conductivity is dominated by the material exhibiting the thermal Hall effect, i.e. $k\bs{I}$.
However, no thermal Hall effect was considered here since $a=0$.
In Case 1-2 and 1-3, the design variable fields were set to $\xi=-1$, $\eta=-1$, $s=0$, and $a=1$, and $\xi=-1$, $\eta=-1$, $s=0$, and $a=-1$, respectively.
The thermal conductivity tensors of the anisotropic material for these cases were the same as in Case 1.
In these cases, the thermal Hall effect was considered for the magnetic field opposite to each other.

Figure~\ref{fig:result0} shows the results of the forward analysis.
These figures indicate that the heat fluxes were curved due to the thermal Hall effect, while the order of temperatures in each case was the same.
This result implies that the design of the asymmetric part of the thermal conductivity tensor will realize the heat flux control, which cannot be achieved only with the symmetric property.

\subsection{Temperature minimization}
%%%%%%% figure %%%%%%%%%%%%%%
\begin{figure}[t]
\centering
\includegraphics[width=6cm]{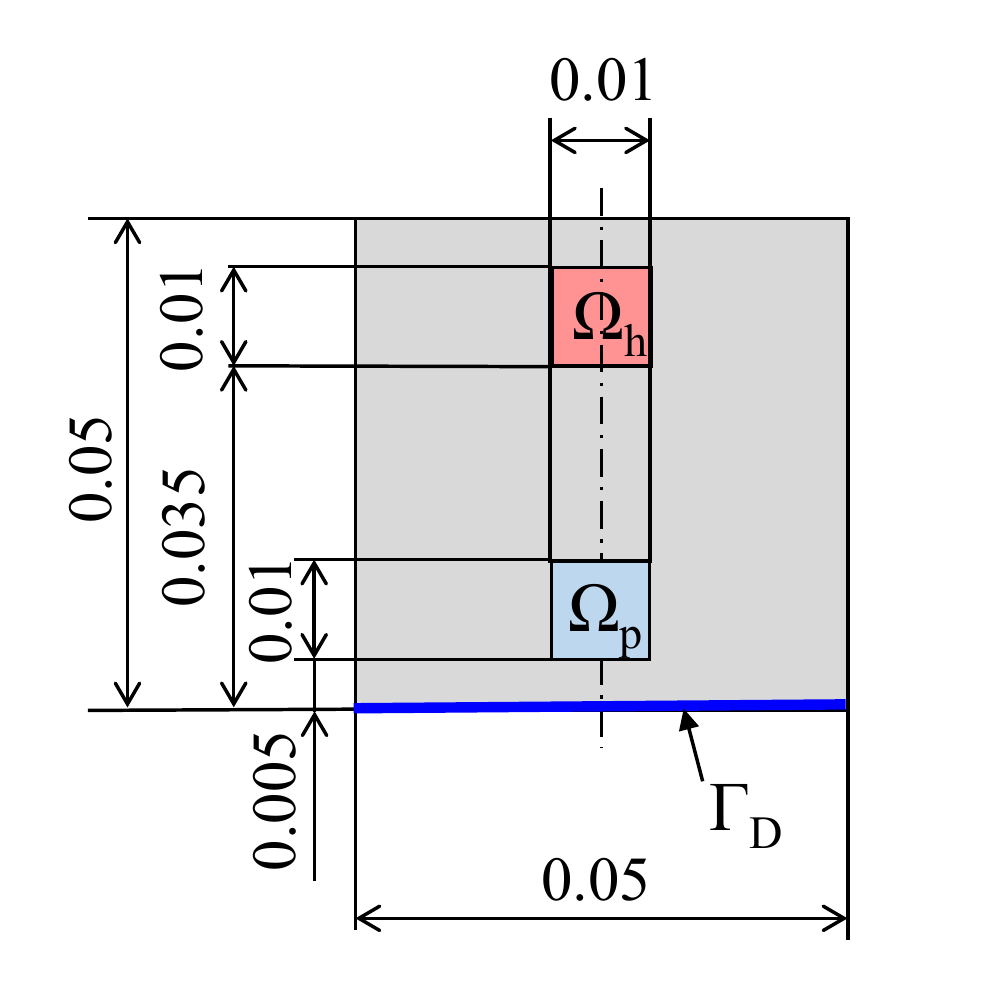}
\caption{Problem settings for the temperature minimization. \label{fig:problem1}}
\end{figure}
%%%%%%% figure %%%%%%%%%%%%%%
%%%%%%% figure %%%%%%%%%%%%%%
\begin{figure*}[t]
\centering
\subfloat[]{\includegraphics[width=0.2\textwidth]{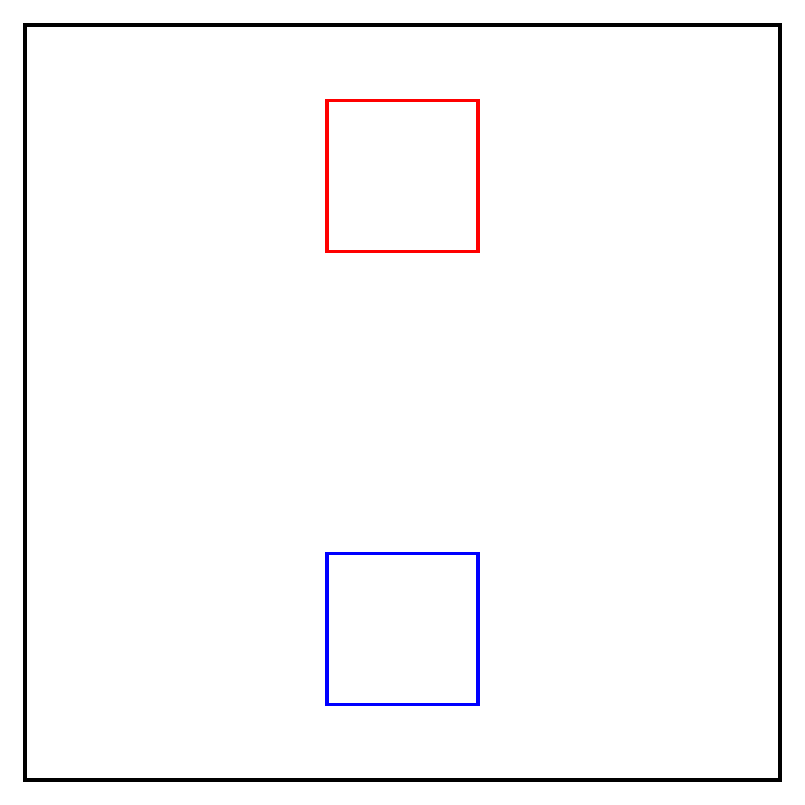} \label{fig:prob1_b0_epsi1_ori}} 
\subfloat[]{\includegraphics[width=0.2\textwidth]{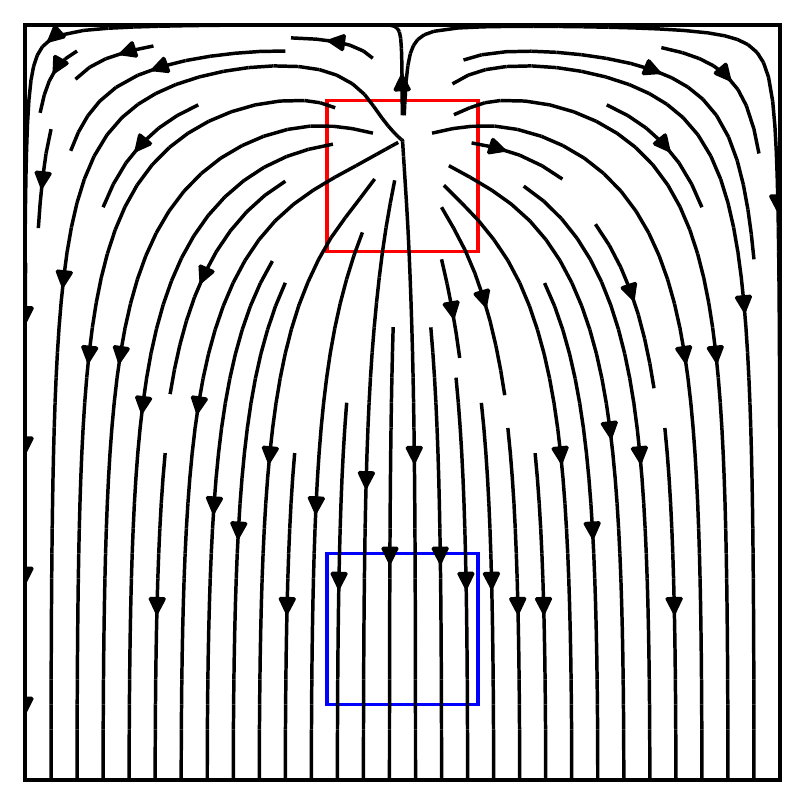} \label{fig:prob1_b0_epsi1_flux}}
\subfloat[]{\includegraphics[width=0.25\textwidth]{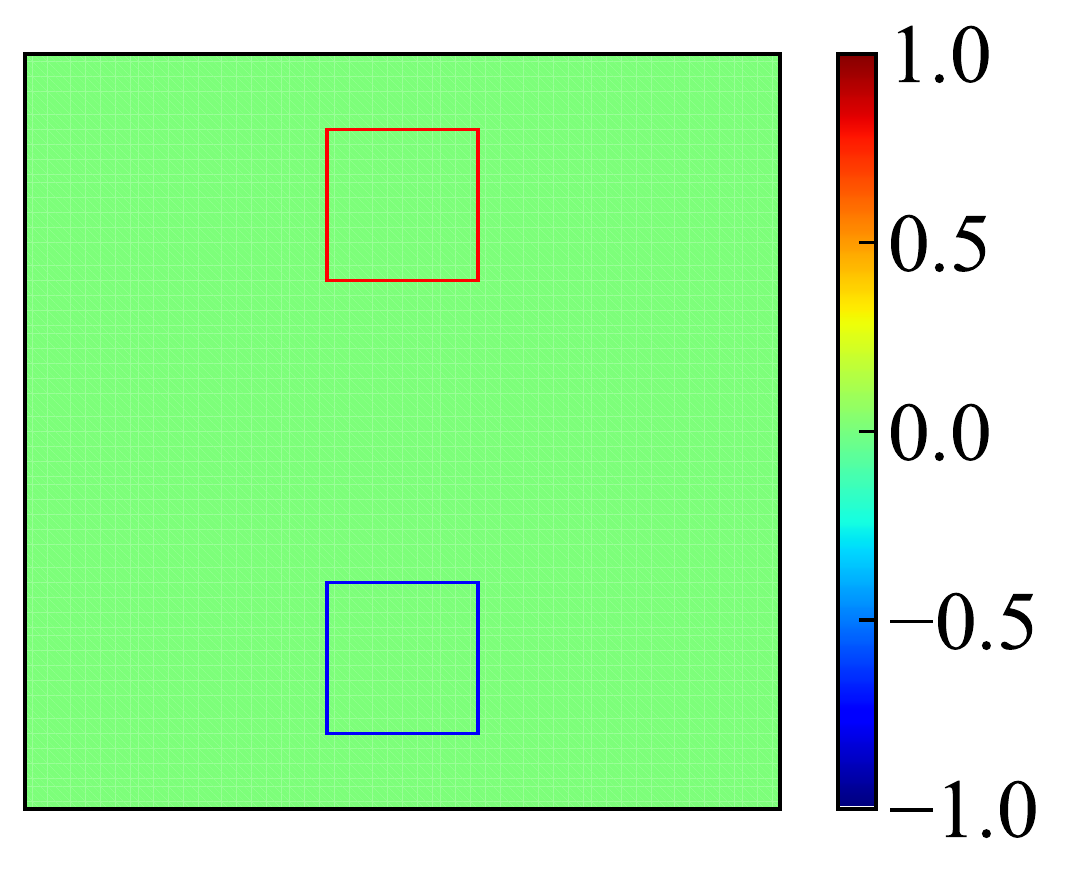} \label{fig:prob1_b0_epsi1_asym}}
\subfloat[]{\includegraphics[width=0.25\textwidth]{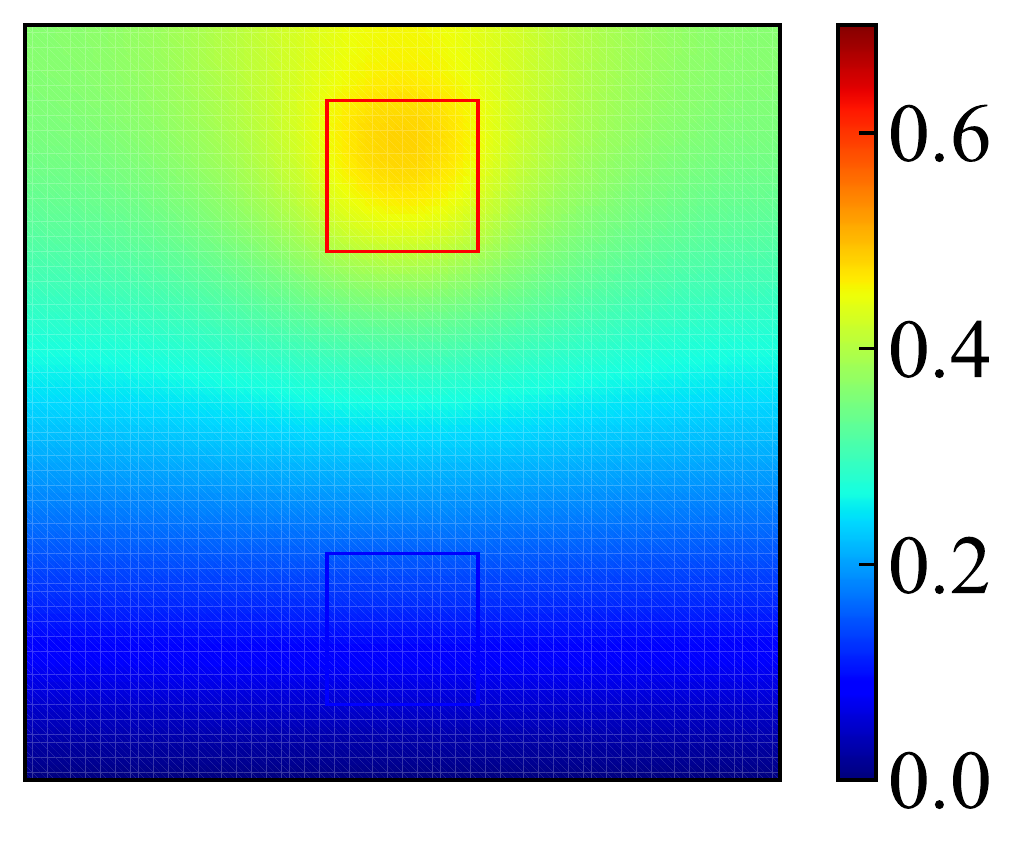} \label{fig:prob1_b0_epsi1_temp}}
\caption{Results of Case 2-1 whose effective thermal conductivity was isotropic and symmetric. (a) The orientation direction. (b) The heat flux. (c) The design variable field $a$. (d) The temperature. The objective function value was $1.02 \times 10^{-5}$. \label{fig:result1-1} }
\end{figure*}
%%%%%%% figure %%%%%%%%%%%%%%
%%%%%%% figure %%%%%%%%%%%%%%
\begin{figure*}[t]
\centering
\subfloat[]{\includegraphics[width=0.2\textwidth]{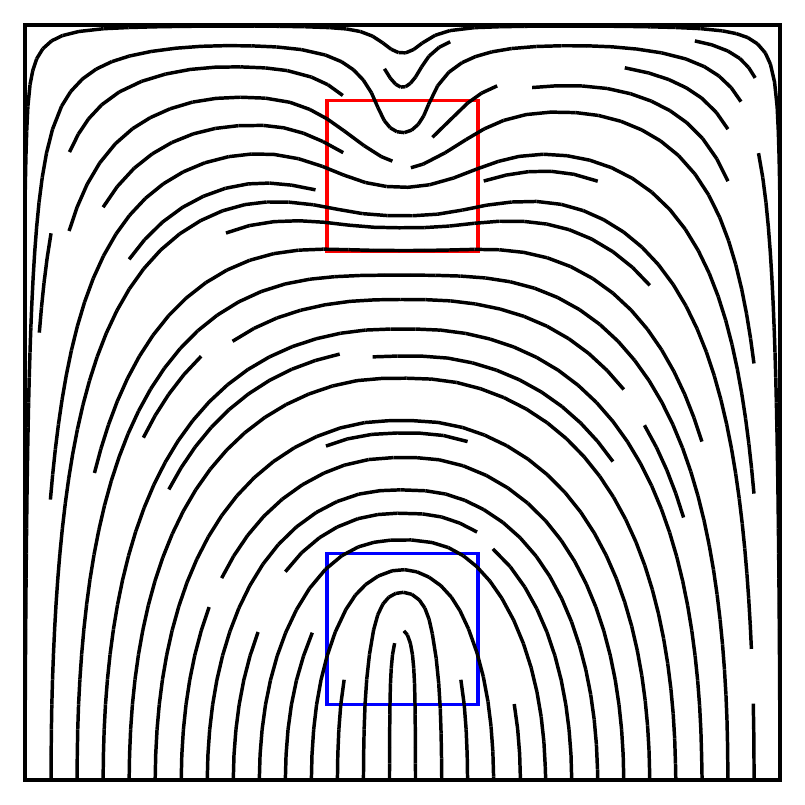} \label{fig:prob1_b0_ori}} 
\subfloat[]{\includegraphics[width=0.2\textwidth]{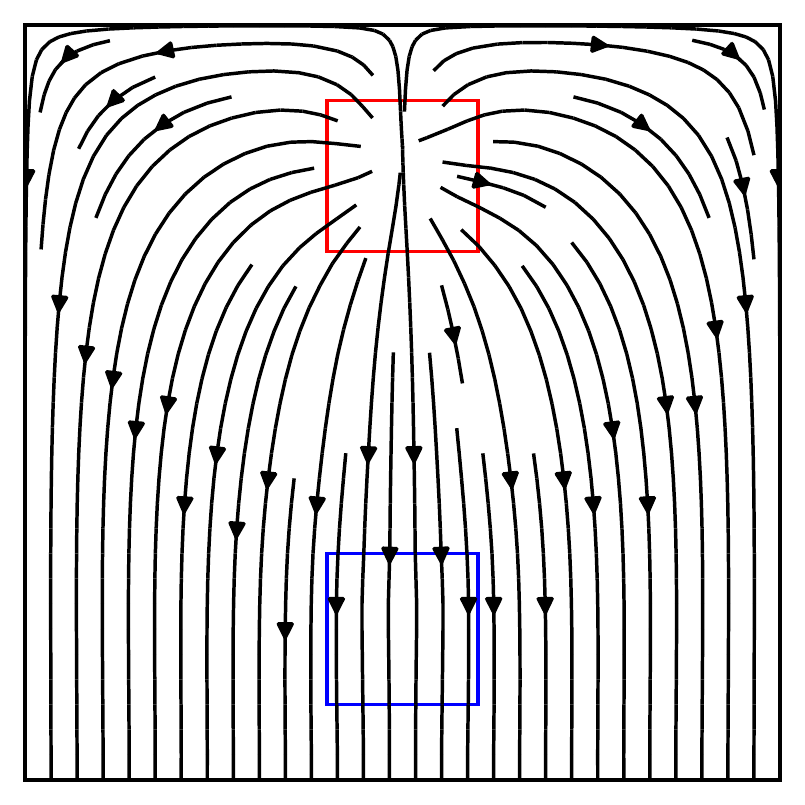} \label{fig:prob1_b0_flux}}
\subfloat[]{\includegraphics[width=0.25\textwidth]{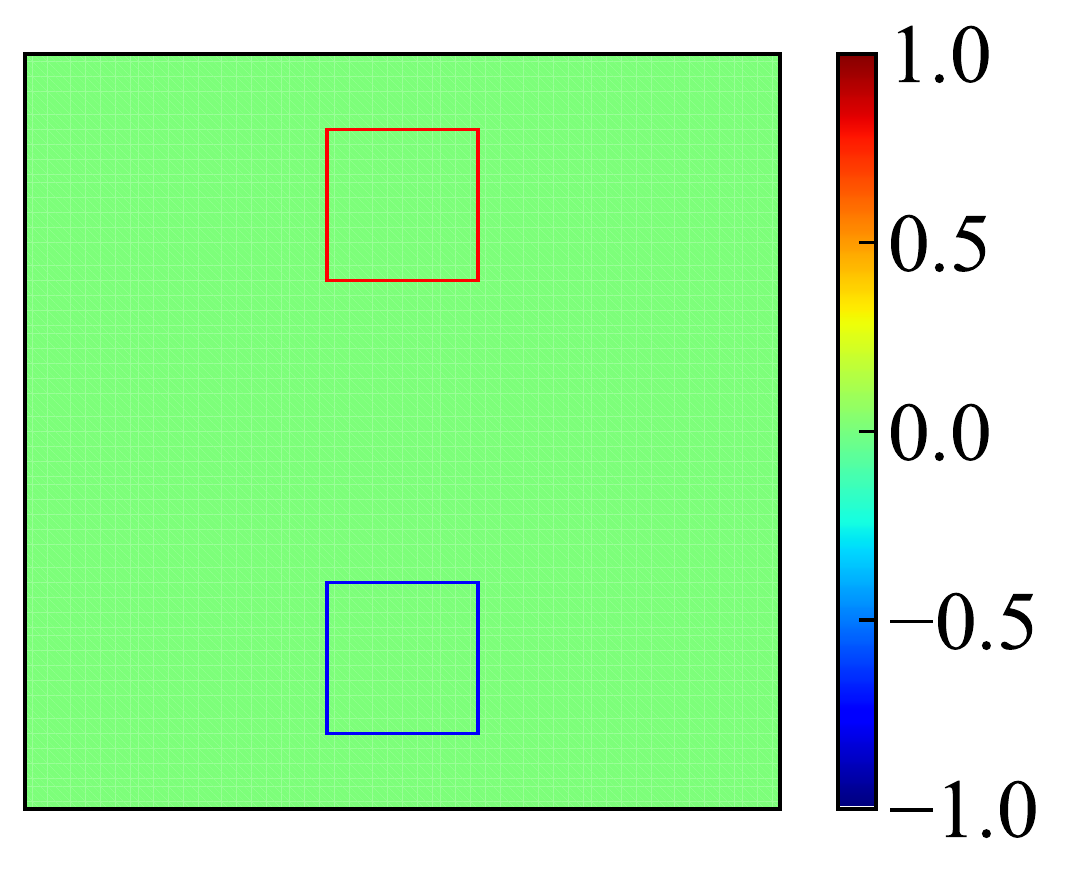} \label{fig:prob1_b0_asym}}
\subfloat[]{\includegraphics[width=0.25\textwidth]{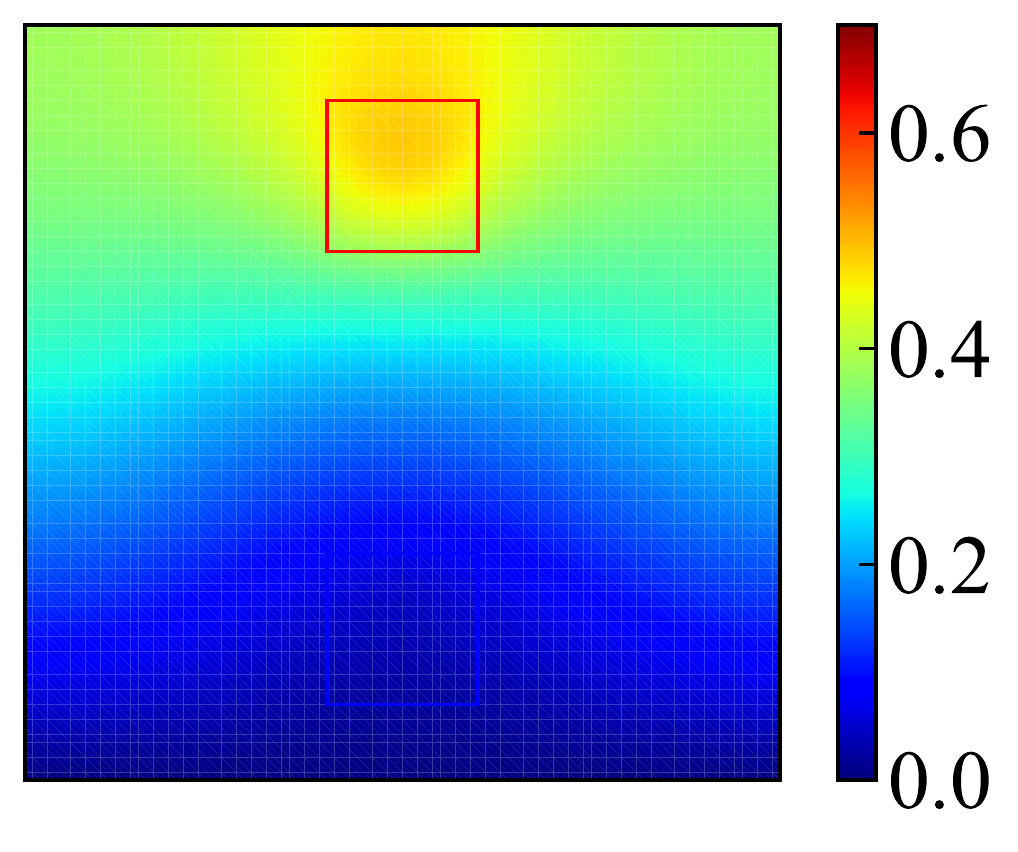} \label{fig:prob1_b0_temp}}
\caption{Optimization results of Case 2-2 whose effective thermal conductivity was anisotropic and symmetric. (a) The orientation direction. (b) The heat flux. (c) The design variable field $a$. (d) The temperature. The objective function value was $4.01 \times 10^{-6}$. \label{fig:result1-2} }
\end{figure*}
%%%%%%% figure %%%%%%%%%%%%%%
%%%%%%% figure %%%%%%%%%%%%%%
\begin{figure*}[t]
\centering
\subfloat[]{\includegraphics[width=0.2\textwidth]{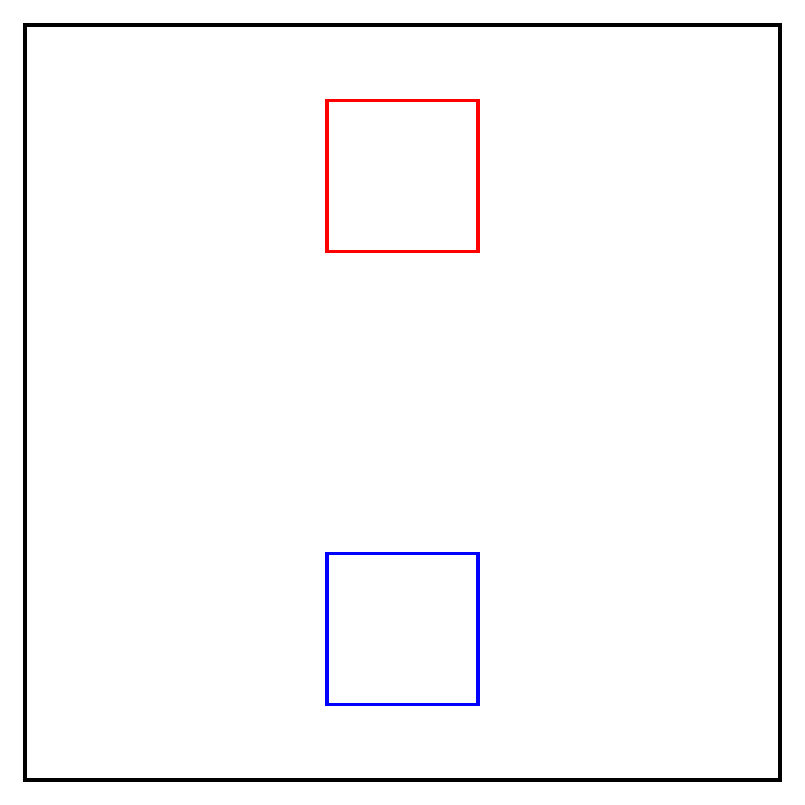} \label{fig:prob1_b03_epsi1_ori}} 
\subfloat[]{\includegraphics[width=0.2\textwidth]{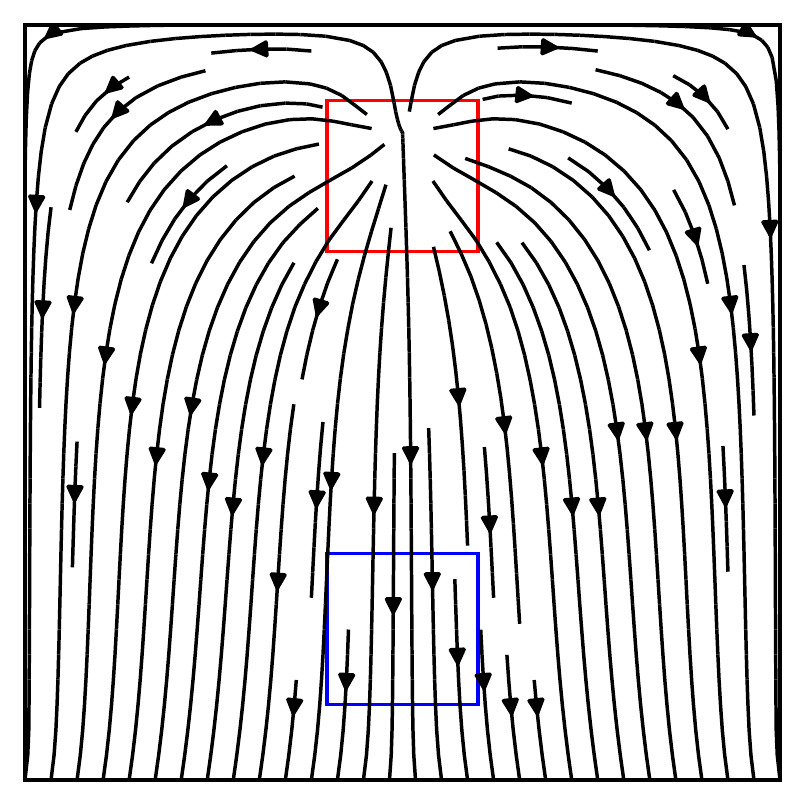} \label{fig:prob1_b03_epsi1_flux}}
\subfloat[]{\includegraphics[width=0.25\textwidth]{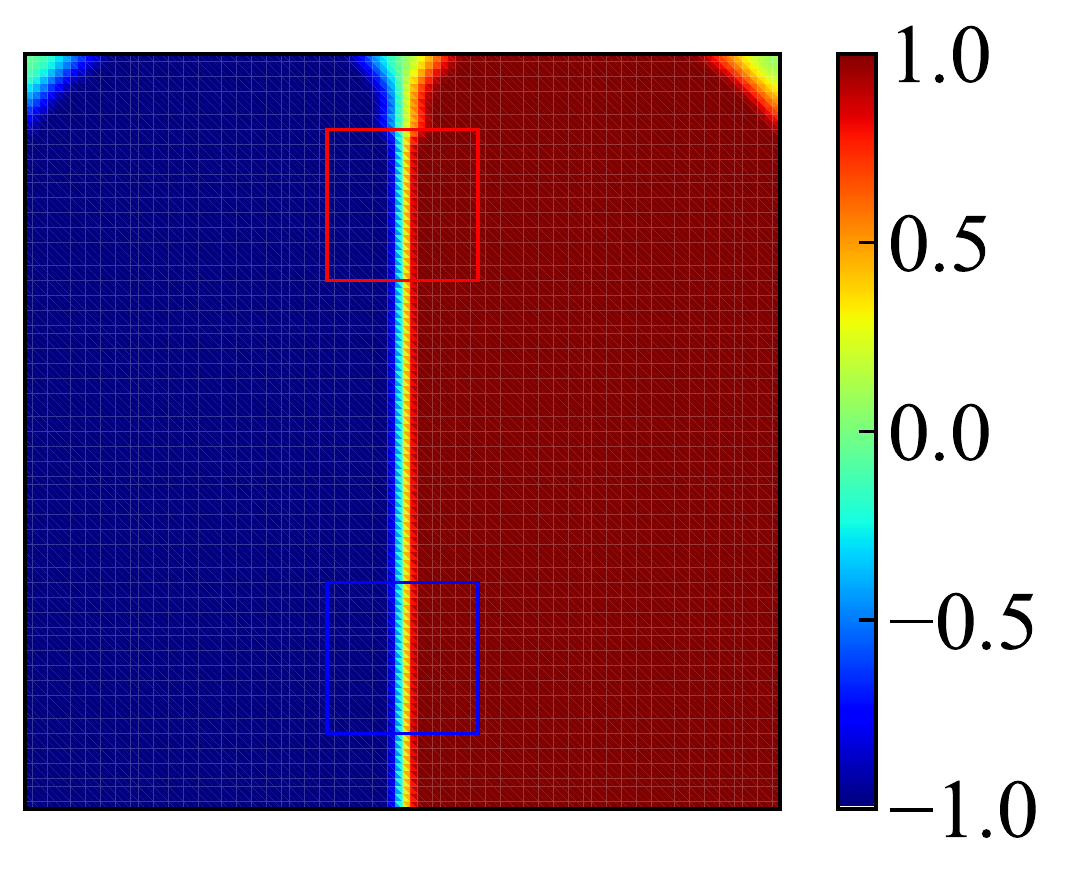} \label{fig:prob1_b03_epsi1_asym}}
\subfloat[]{\includegraphics[width=0.25\textwidth]{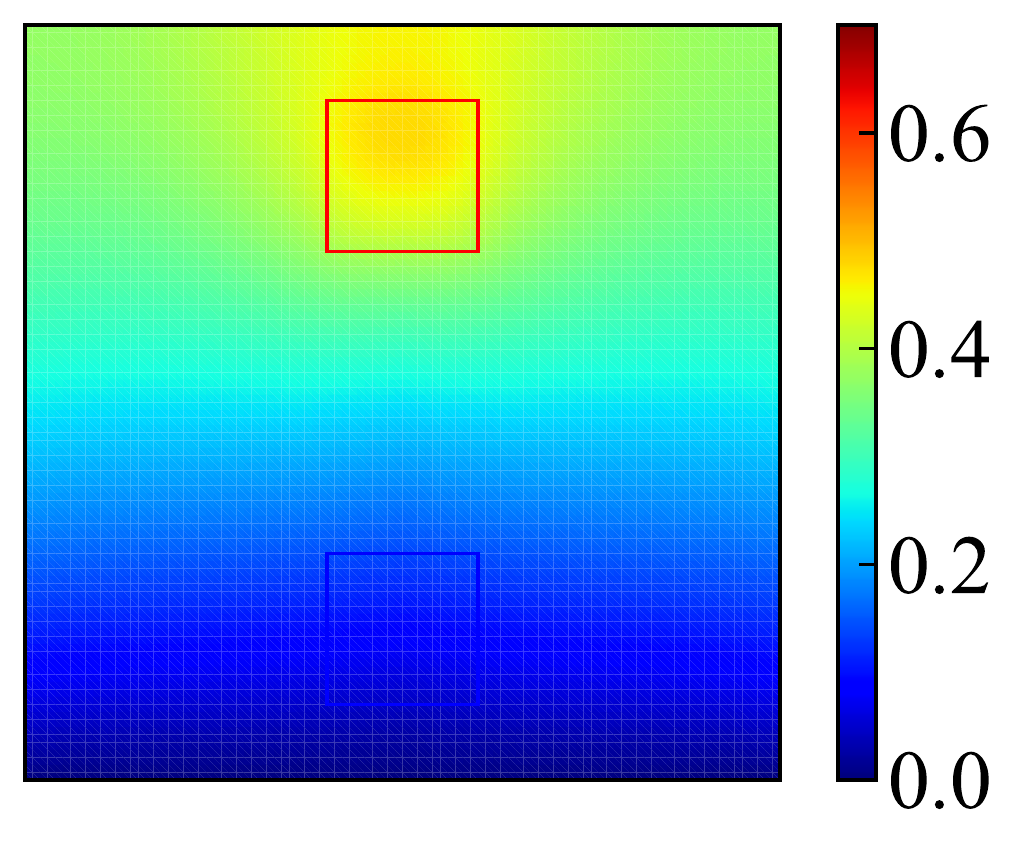} \label{fig:prob1_b03_epsi1_temp}}
\caption{Optimization results of Case 2-3 whose effective thermal conductivity was isotropic and asymmetric. (a) The orientation direction. (b) The heat flux. (c) The design variable field $a$. (d) The temperature. The objective function value was $9.05 \times 10^{-6}$. \label{fig:result1-3} }
\end{figure*}
%%%%%%% figure %%%%%%%%%%%%%%
%%%%%%% figure %%%%%%%%%%%%%%
\begin{figure*}[t]
\centering
\subfloat[]{\includegraphics[width=0.2\textwidth]{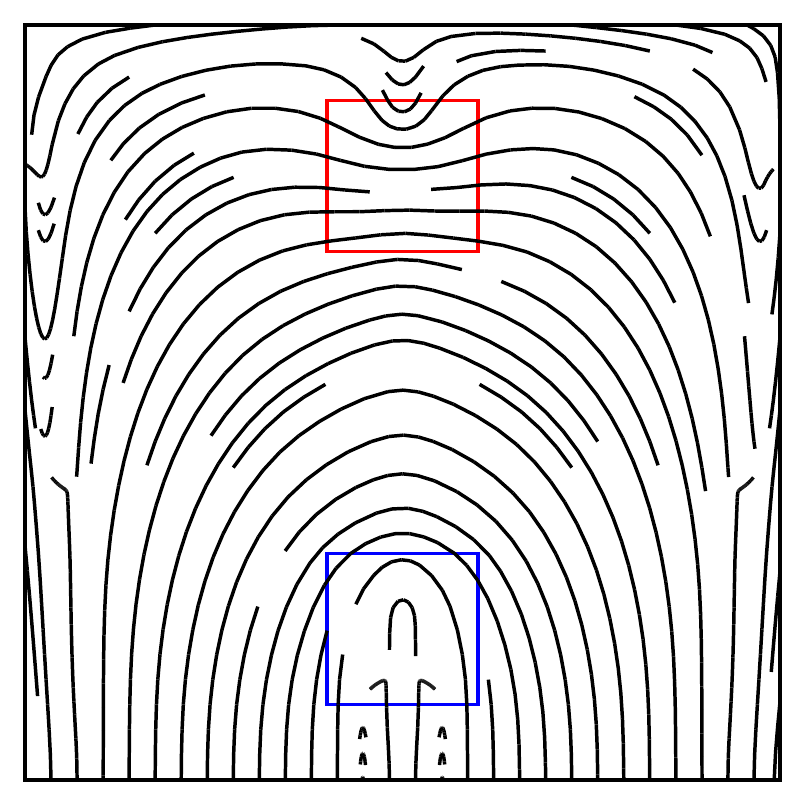} \label{fig:prob1_b03_ori}} 
\subfloat[]{\includegraphics[width=0.2\textwidth]{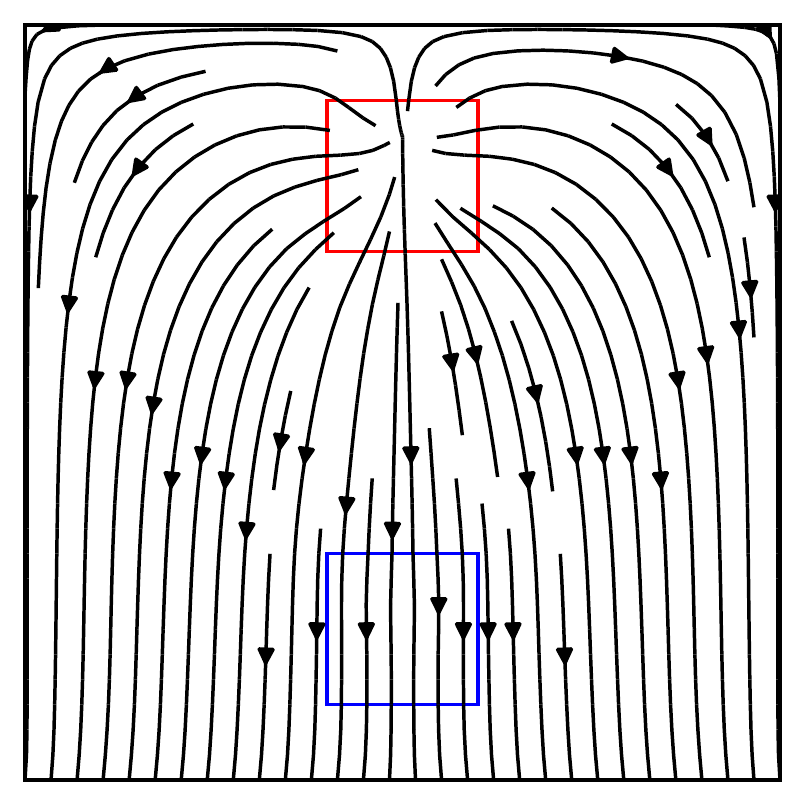} \label{fig:prob1_b03_flux}}
\subfloat[]{\includegraphics[width=0.25\textwidth]{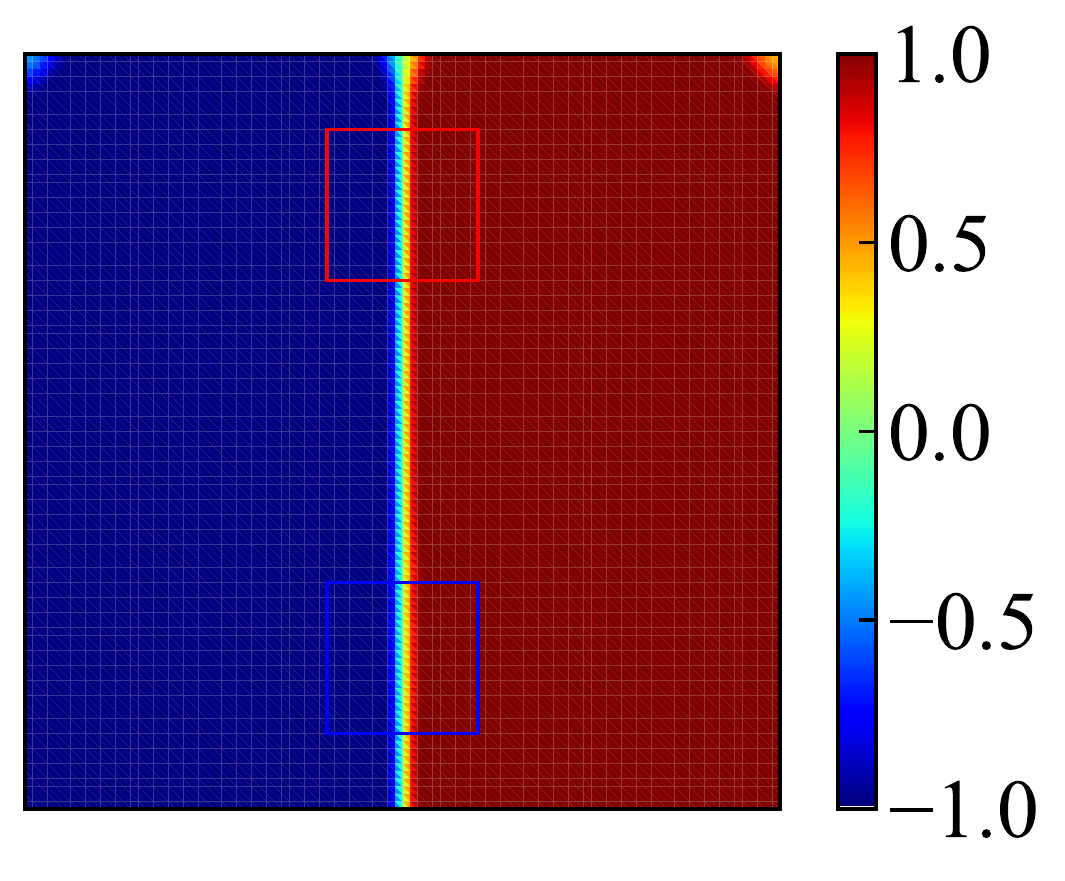} \label{fig:prob1_b03_asym}}
\subfloat[]{\includegraphics[width=0.25\textwidth]{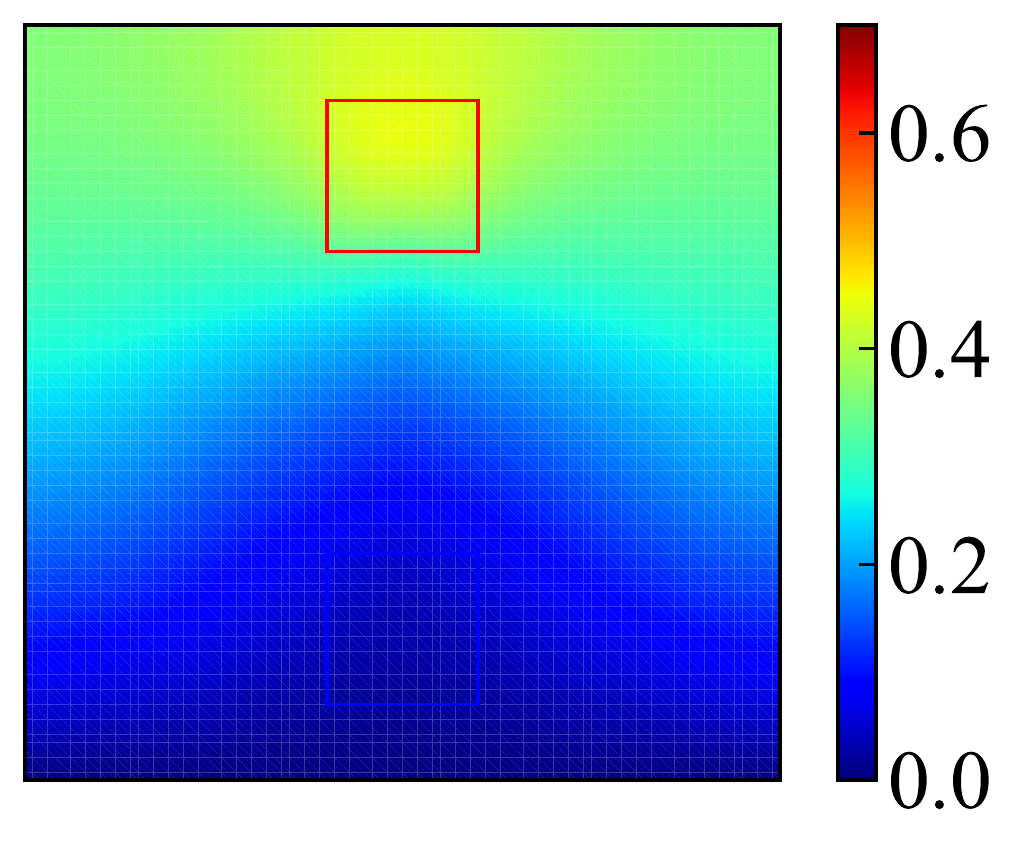} \label{fig:prob1_b03_temp}}
\caption{Optimization results of Case 2-4 whose effective thermal conductivity was anisotropic and asymmetric. (a) The orientation direction. (b) The heat flux. (c) The design variable field $a$. (d) The temperature. The objective function value was $3.30 \times 10^{-6}$. \label{fig:result1-4} }
\end{figure*}
%%%%%%% figure %%%%%%%%%%%%%%
Next, we examined the effectiveness of the asymmetric part of the thermal conductivity tensor through the temperature minimization problem.
Figure~\ref{fig:problem1} illustrates the problem settings.
We considered the following four cases, based on the property of the effective thermal conductivity:
\begin{enumerate}
\setlength{\leftskip}{12mm}
    \item[Case 2-1] Isotropic and symmetric
    \item[Case 2-2] Anisotropic and symmetric
    \item[Case 2-3] Isotropic and asymmetric
    \item[Case 2-4] Anisotropic and asymmetric
\end{enumerate}
The isotropic material was prepared by setting $\varepsilon = \varepsilon' = 1.0$, which resulted in $k_{11} = k_{22} = c/2$ and $k_{12} = 0$.

Figure~\ref{fig:result1-1} shows the results of Case 2-1, where the effective thermal conductivity tensor was isotropic and symmetric.
In this case, there is no room for optimization because $b=0$ and $\varepsilon = \varepsilon' = 1.0$.
The objective function value was $J=1.02 \times 10^{-5}$.

Figure~\ref{fig:result1-2} shows the optimization results of Case 2-2 where the thermal conductivity tensor was restricted to the symmetric but can be anisotropic.
The orientation direction of the anisotropic material shown in Fig.~\ref{fig:result1-2}\subref{fig:prob1_b0_ori} was aligned so that the heat flowed avoiding the domain $\Omega_\mathrm{p}$.
As a result, the temperature in the domain became smaller than that of the isotropic material as shown in Figs.~\ref{fig:result1-1}\subref{fig:prob1_b0_epsi1_temp} and \ref{fig:result1-2}\subref{fig:prob1_b0_temp}.
The objective function value was $J=4.01 \times 10^{-6}$.

Figure~\ref{fig:result1-3} shows the optimization results of Case 2-3 where the effective thermal conductivity tensor was isotropic and asymmetric.
Compared with the previous two cases: Cases 2-1 and 2-2, the heat flux was slightly curved to avoid the domain $\Omega_\mathrm{p}$, which resulted in the low temperature as shown in Fig.~\ref{fig:result1-3}\subref{fig:prob1_b03_epsi1_temp}.
This heat-avoidance effect was realized by having opposite values of $a$ in the left and right half domains as shown in Fig.~\ref{fig:result1-3}\subref{fig:prob1_b03_epsi1_asym}.
The objective function value was $J=9.05 \times 10^{-6}$, which was smaller than that of Case 2-1 where the thermal conductivity tensor had the isotropic and symmetric property, but was larger than that of Case 2-2 where the tensor was anisotropic and symmetric.
This result implies that the anisotropic property is more effective to minimize the temperature than the asymmetric property when $b=0.3$ which represents the relatively large thermal Hall effect.

Figure~\ref{fig:result1-4} shows the optimization results where the effective thermal conductivity tensor was anisoropic and asymmetric.
The orientation direction of the anisotropic material shown in Fig.~\ref{fig:result1-4}\subref{fig:prob1_b03_ori} was similar to that shown in Fig.~\ref{fig:result1-2}\subref{fig:prob1_b0_ori}.
That is, the orientation direction of the anisotropic material was aligned so that the heat flowed avoiding the domain $\Omega_\mathrm{p}$.
The asymmetric part of the thermal conductivity tensor $a$ of this case, shown in Fig.~\ref{fig:result1-4}\subref{fig:prob1_b03_asym} was similar to that of Case 2-3 shown in Fig.~\ref{fig:result1-3}\subref{fig:prob1_b03_epsi1_asym}.
That is, the result of this case had both characteristics of the results of Cases 2-2 and 2-3.
As a result, this case had the smallest objective function value of $J=3.30 \times 10^{-6}$.

%%%%%%% figure %%%%%%%%%%%%%%
\begin{figure}[t]
\centering
\includegraphics[width=6cm]{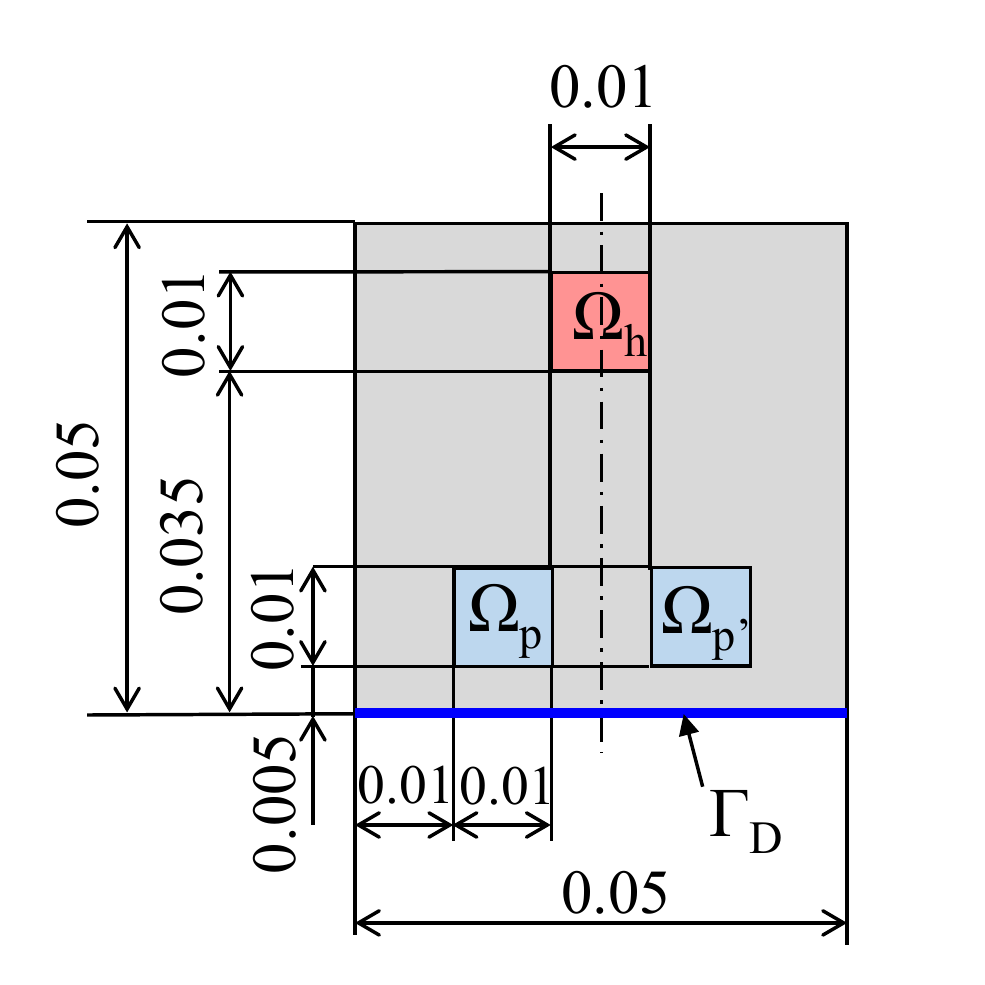}
\caption{Problem settings for the heat path switching problem. \label{fig:problem2}}
\end{figure}
%%%%%%% figure %%%%%%%%%%%%%%
%%%%%%% figure %%%%%%%%%%%%%%
\begin{figure*}[t]
\centering
\subfloat[]{\includegraphics[width=0.2\textwidth]{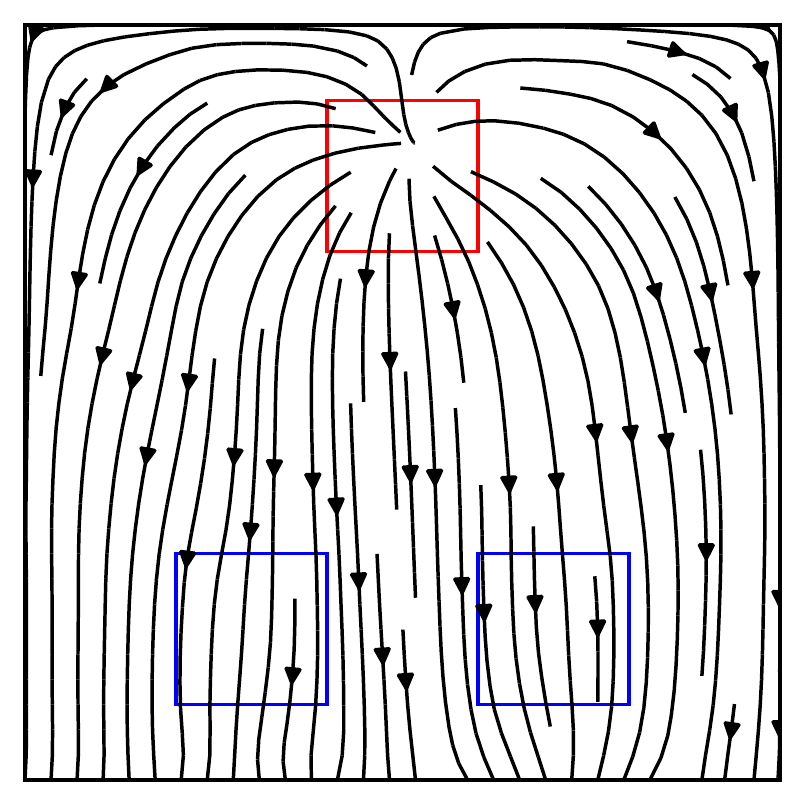} \label{fig:prob2_b03_flux1}} 
\subfloat[]{\includegraphics[width=0.25\textwidth]{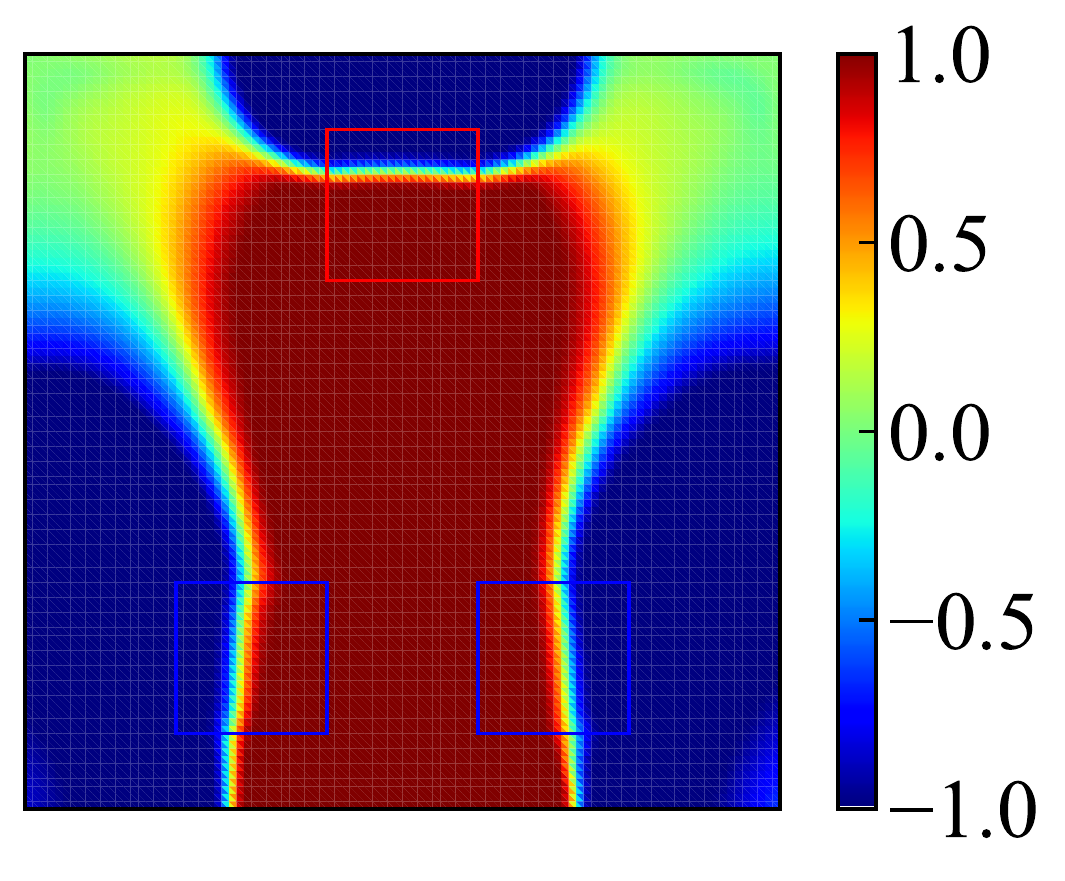} \label{fig:prob2_b03_asym1}}
\subfloat[]{\includegraphics[width=0.25\textwidth]{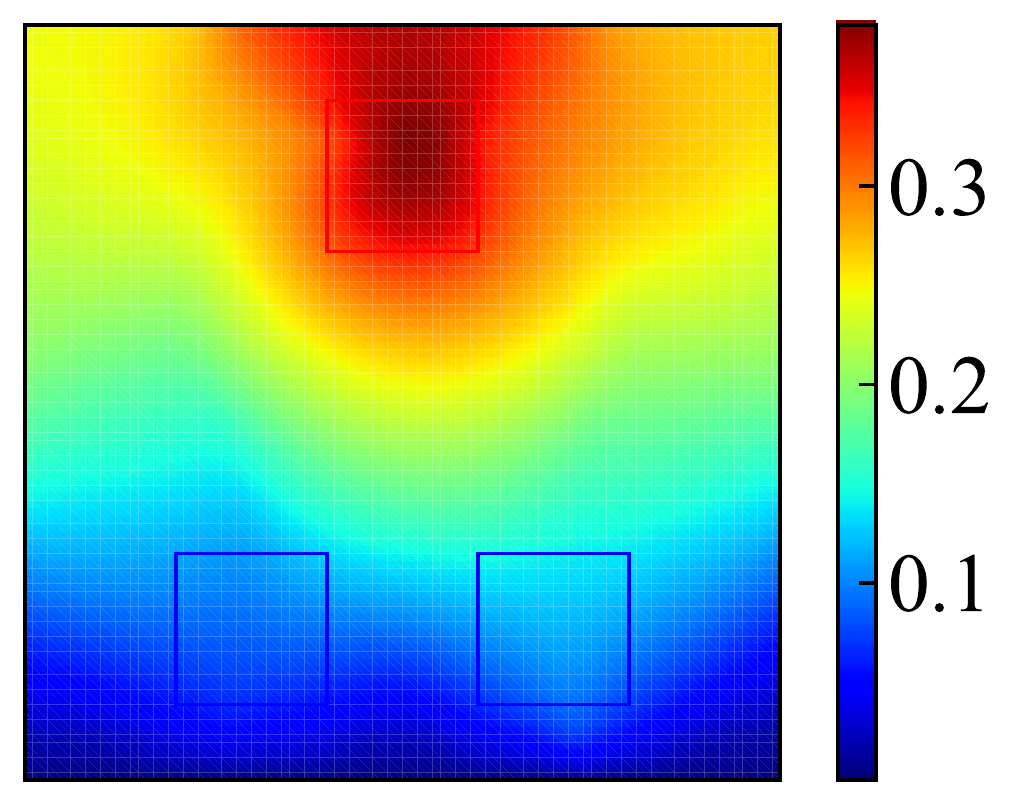} \label{fig:prob2_b03_temp1}} \\
\subfloat[]{\includegraphics[width=0.2\textwidth]{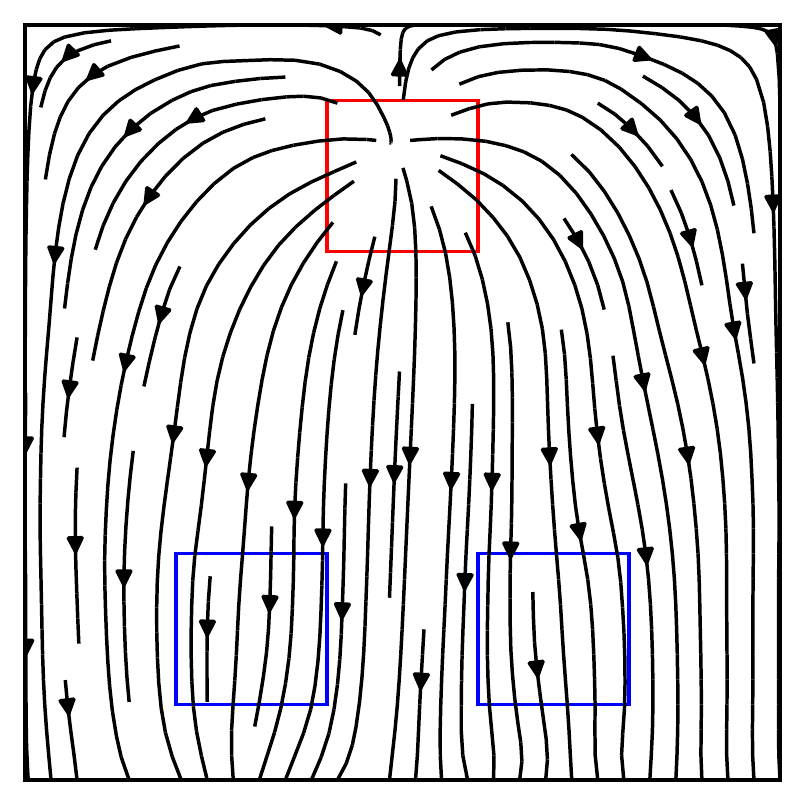} \label{fig:prob2_b03_flux2}} 
\subfloat[]{\includegraphics[width=0.25\textwidth]{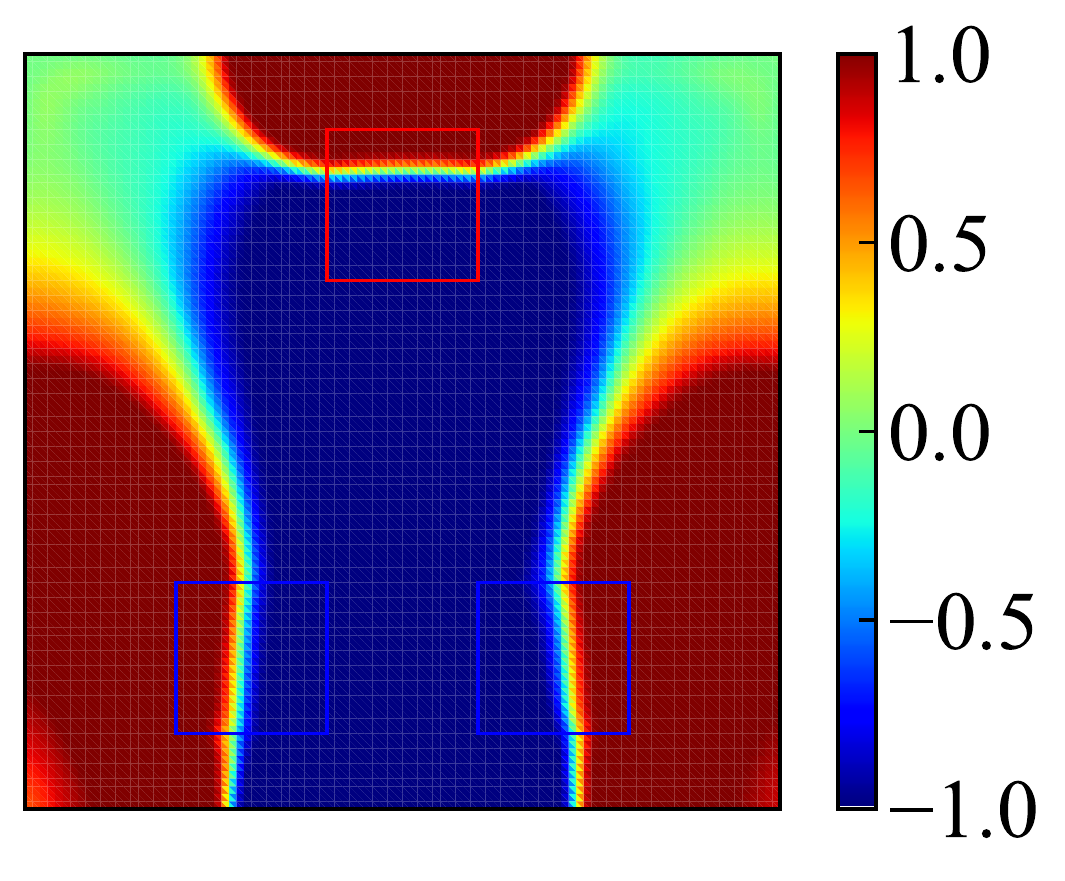} \label{fig:prob2_b03_asym2}}
\subfloat[]{\includegraphics[width=0.25\textwidth]{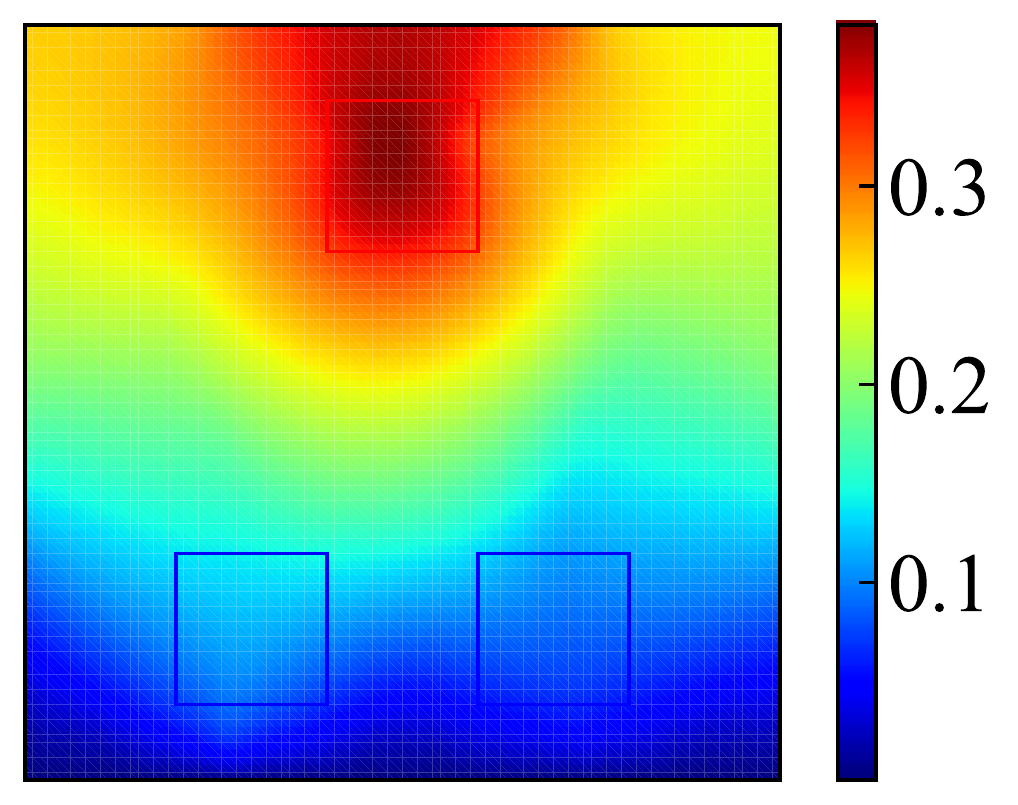} \label{fig:prob2_b03_temp2}} \\
\subfloat[]{\includegraphics[width=0.2\textwidth]{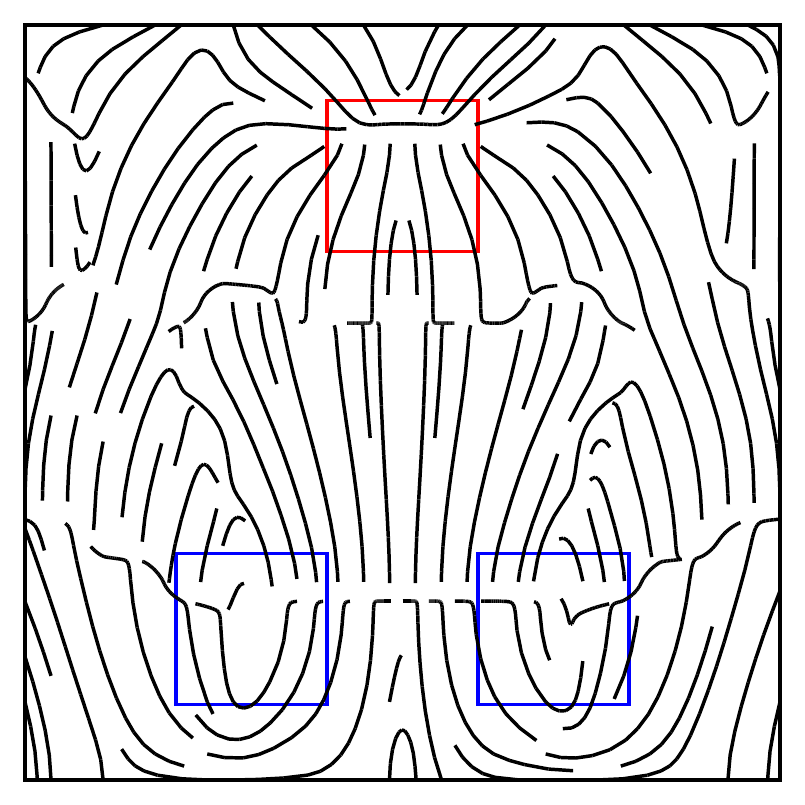} \label{fig:prob2_b03_ori}} 
\caption{Optimization results of the heat path switching problem. (a) The heat flux of Mode 1. (b) The design variable field $a$. (c) The temperature of Mode 1. (d) The heat flux of Mode 2. (e) The design variable field $a'$. (f) the temperature of Mode 2. Mode 1 represents the analysis using the asymmetric part of the effective thermal conductivity tensor determined by the design variable field $a$, while Mode 2 represents it determined by $a'$. \label{fig:result2} }
\end{figure*}
%%%%%%% figure %%%%%%%%%%%%%%

\subsection{Heat path switching problem}
Finally, we examined the effectiveness of the proposed method through the heat path switching problem, which could not be solved without the asymmetric part of the thermal conductivity tensor.

Figure~\ref{fig:problem2} illustrates the problem settings.
Here, we set parameters for ensuring the positiveness of $\bs{k}^\text{aniso}$ to $\varepsilon = 1.0 \times 10^{-4}$ and $\varepsilon' = 1.0 \times 10^{-4}$, respectively.

Here, we considered two kinds of the effective thermal conductivity tensor with different asymmetric parts by introducing two design variable fields, respectively denoted by $a$ and $a'$ as described in Section~\ref{sec:objective}.
These tensor settings correspond to two kinds of magnetic fields applied to the material.
We call the analysis using the design variable field $a$
Mode 1, and call the analysis using $a'$ Mode 2.

Figure~\ref{fig:result2} shows the optimization result of the heat path switching problem.
As shown in Figs.~\ref{fig:result2}\subref{fig:prob2_b03_flux1} and \subref{fig:prob2_b03_flux2}, the heat fluxes were curved according to the asymmetric properties shown in Figs.~\ref{fig:result2}\subref{fig:prob2_b03_asym1} and \subref{fig:prob2_b03_asym2}, which had opposite signs to each other.
As a result, the temperature differences between domains $\Omega_\mathrm{p}$ and $\Omega_\mathrm{p'}$ arose depending on the modes as illustrated in Figs.~\ref{fig:result2}\subref{fig:prob2_b03_temp1} and \subref{fig:prob2_b03_temp2}.
Interestingly, the orientation direction shown in Fig.~\ref{fig:result2}\subref{fig:prob2_b03_ori} was aligned to prevent the heat from flowing near the lower boundary $\Gamma_\text{D}$, which was a different trend from the result of the temperature minimization problem.
This result indicates that the thermal Hall effect was used as a gate to make it easier for the heat to flow through the material that initially had the thermal insulation property.
This experiment demonstrates that our proposed method can appropriately deal with the asymmetric constitutive tensor for controlling heat flows.

\section{Conclusions} \label{sec:conclusion}
This paper presented free material optimization (FMO) for asymmetric thermal conductivity tensors.
The following are the summary of this paper.
\begin{enumerate}
    \item We focused on the thermal Hall effect to control heat flows and modeled it in the asymmetric part of the thermal conductivity tensor. Combining the anisotropic property of the tensor, we parametrized the effective thermal conductivity tensor by introducing four design variable fields so that the physical requirements of the constitutive tensor could be automatically satisfied.
    \item We constructed the optimization algorithm, a combination of the optimizer based on the reaction-diffusion equation and the adaptive moment estimation. The design variable fields were smoothed due to the diffusion term, and the use of the first and the second moment of gradients based on the adaptive moment estimation enabled the optimization process to be stable.
    \item Several numerical experiments were provided to demonstrate that our proposed method can appropriately deal with the asymmetric constitutive tensor for controlling heat flows. The second group of experiments implied that the thermal Hall effect helped the temperature minimization, whereas the anisotropic property was more effective in achieving the temperature minimization than the asymmetric property. The third group of experiments showed heat path switching using the thermal Hall effect, which could not be realized without the asymmetric part of the thermal conductivity tensor.
\end{enumerate}

\backmatter

\section*{Statements and Declarations}
\bmhead{Conflict of interest}
On behalf of all authors, the corresponding author states that there is no conflict of interest.
\bmhead{Replication of results}
The source code is unavailable due to institutional constraints. However, further algorithm details are available upon request to the authors.

\bibliography{sn-bibliography}% common bib file

\end{document}